\documentclass[12pt,aps,tightenlines,notitlepage,superscriptaddress]{revtex4-1}

\usepackage{amsmath}
\usepackage{amsthm}
\usepackage{amsfonts}
\usepackage{amssymb}
\usepackage{bm}							
\usepackage{enumerate}					
\usepackage{graphicx}					
\usepackage{mathtools}                  
\usepackage{tikz}

\newcommand{\mathd}{\mathrm{d}}
\newcommand{\mathe}{\mathrm{e}}

\newcommand{\myhbar}{\overline{h}}

\newcommand{\surften}{\gamma}

\newcommand{\Lagrange}{\mathcal{L}}
\newcommand{\Energy}{\mathcal{E}}
\newcommand{\lagrange}{\ell}

\newcommand{\lie}{\pounds}

\newcommand{\thetaeq}{\theta_{\mathrm{eq}}}

\newcommand{\argmin}{\operatornamewithlimits{argmin}}

\graphicspath{{figures/}}

\begin{document}

\title{A mathematical model and  mesh-free numerical method for contact-line motion in lubrication theory}
\author{Khang Ee Pang, Lennon \'O N\'araigh}
\affiliation{School of Mathematics and Statistics, University College Dublin}
\date{\today}
\begin{abstract}
We introduce a mathematical model with a mesh-free numerical method to describe contact-line motion in lubrication theory.  We show how the model  resolves the singularity at the contact line, and generates smooth profiles for an evolving, spreading droplet.    The model describes well the physics of droplet spreading -- including Tanner's Law for the evolution of the contact line.  The model can be configured to describe complete wetting or partial wetting, and we explore both cases numerically.  In the case of partial wetting, the model also admits analytical solutions for the droplet profile, which we present here.  
\end{abstract}
\maketitle

\section{Introduction}

When a droplet of fluid (surrounded by a gaseous atmosphere) is deposited on a substrate, it spreads until it reaches an equilibrium configuration.  At equilibrium, the angle between the liquid-gas interface and the solid surface is measured (conventionally, through the liquid), to yield the equilibrium contact angle $\thetaeq$. If the angle is less than $90^\circ$, the substrate is deemed hydrophilic, whereas if the angle exceeds $90^\circ$, the substrate is deemed hydrophobic~\cite{de1985wetting}.  Droplet spreading then describes the dynamic phase before the attainment of this equilibrium.

In droplet spreading, the point of contact between the substrate and the gas-liquid interface is in motion.  And yet this contradicts the classical no-slip assumption in viscous fluid flow, which stipulates that there should be no relative motion between a substrate in contact with a fluid~\cite{huh1971hydrodynamic, dussan1974on}.  The resolution of this paradox is that there is missing physics, and that on a sufficiently small scale, there is slip, the dynamics of which are governed by the interactions between the fluid molecules and the substrate molecules~\cite{dussan1979on}.  These molecular-level interactions can be incorporated into a macroscopic fluid model via a so-called regularization technique.  Broadly, there are three regularization techniques in the literature -- the slip length~\cite{hocking1981sliding}, the precursor film~\cite{de1985wetting}, and the diffuse interface~\cite{pismen2000disjoining,ding2007wetting}.  Although methodologically distinct, these yield the same qualitative and quantitative results when used to model droplet dynamics.  This consistency between the different approaches gives a solid justification for the general approach of model regularization.

The purpose of this article is to introduce a novel model regularization, albeit one in the spirit of those just described.  The focus of the work is on mode regularization for the case of thin-film flows (also called lubrication flows).  For simplicity, we focus on two-dimensional configurations (or equivalently, three-dimensional axi-symmetric configurations), however, the generalization to three dimensions is straightforward.
%
%
%
%
%
%
%
%
%
\begin{figure}
	\centering
		\includegraphics[width=1\textwidth]{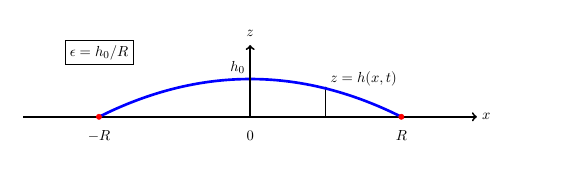}
		\caption{Schematic description of droplet spreading on a substrate}
	\label{fig:sketch1}
\end{figure}

Thin-film flows arise in the context of hydrophilic substrates, where the equilibrium shape of the droplet is such that the typical  size of the droplet base $R$ greatly exceeds the maximum droplet height $h_0$, leading to a small parameter $\epsilon=h_0/R$, with $\epsilon\ll 1$.  In this context, the Navier--Stokes equations reduce down to a single equation for the interface height (the so-called Thin-Film Equation).   In this context also, the interface height is conventionally written as $z=h(x,t)$.  Thus, the coordinate $x$ and is in the plane of the substrate, and the $z$-coordinate is orthogonal to the substrate (e.g. Figure~\ref{fig:sketch1}).


\subsection*{Motivation for this work}

In the context of droplet spreading, the Thin-Film Equation inherits the contact-line singularity from the full Navier--Stokes equations; the singularity can be regularized  by introducing a slip length
~\cite{hocking1981sliding,sibley2015asymptotics}, or a precursor film~\cite{de1985wetting,bonn2009wetting}.  Both these regularizations give accurate and consistent descriptions of contact-line spreading~\cite{savva2011dynamics}, albeit with some drawbacks -- the classical Navier slip-length model has a logarithmic stress singularity at the contact line, while as the precursor-film model requires a precursor film to be present that extends indefinitely beyond the droplet core.  Although physically, such a precursor film does exist, it has a very small thickness ($10-100\,\mathrm{nm}$~\cite{bonn2009wetting}), meaning that such a small scale must be resolved in the model: in particular, the numerical grid size must be at least as small as the precursor-film thickness~\cite{schwartz1998simulation}.  Furthermore, the resulting equations are quite stiff numerically~\cite{diez2000global}.  Although this approach is just about feasible for millimetre-scale droplets, it may not be feasible for larger ones.  Beyond the millimetre scale, an unphysically large precursor-film thickness can be used in numerical investigations (and the results checked for robustness to changes in the value of the precursor-film thickness), however, this approach is somewhat unsatisfactory.
%
%
%
%
%
%
The Diffuse Interface Model has been proposed as a more general regularization of the contact-line singularity problem, valid for the full Navier--Stokes equations beyond the lubrication limit~\cite{ding2007inertial}.  The Diffuse Interface Model has been implemented for droplet spreading in the thin-film lubrication approximation~\cite{pismen2000disjoining}.

Our contribution in this article is to introduce a novel regularization technique similar in spirit to the Diffuse Interface Method -- we formulate a theory of droplet spreading involving a smooth interface height $\myhbar$, as well as a sharp interface height $h$, which interact via a convolution operator and an evolution equation. 
%
%
%
%
In a previous work~\cite{holm2020gdim}, the idea behind this regularization was introduced in the context of thin-film lubrication flows -- the so-called Geometric Thin Film Equation.  In the present article, we extend this previous work by introducing a particle method as a novel and highly accurate solution method for the Geometric Thin Film Equation.  Also, Reference~\cite{holm2020gdim} was for complete wetting -- in this work we extend the Geometric Thin-Film Equation to describe partial wetting as well.

The Geometric Thin-Film Equation can be viewed as special case of a  mechanical model for energy-dissipation on general configuration spaces -- the derivation of the general model involves methods from Geometric Mechanics such as Lie Derivatives and Momentum Maps~\cite{holm2008formation} -- hence the name \textit{Geometric} Thin-Film Equation.  
The main advantage of this new method so far has been the non-stiff nature of the differential equations in the model, which leads to robust numerical simulation results.  A second advantage (the main focus of the present work) is that the Geometric Thin-Film equation admits so-called particle solutions.  These give rise to an efficient and accurate numerical method (the particle method) for solving the model equations.  


The basis for the particle method lies in  the structure of the Geometric Thin-Film Equation: the model includes a `smoothened' free-surface height $\myhbar(x,t)$ and a `sharp' free-surface height $h(x,t)$, related by convolution, $\myhbar(x,t)=\Phi*h(x,t)$, where $\Phi\geq 0$ is a filter function with a characteristic lengthscale $\alpha$.  The model admits a delta-function solution for the sharp free-surface height $h(x,t)=\sum_{i=1}^N w_i \delta(x-x_i)$, where $\delta$ is the Dirac delta function, and $x_i(t)$ is the (time-dependent) centre of the Delta function.
  The delta-function centres $\{x_i(t)\}_{i=1}^N$ satisfy a set of ordinary differential equations:
\begin{equation}
\frac{\mathd x_i}{\mathd t}=V_i(x_1,\cdots,x_N),\qquad i\in \{1,2,\cdots\,N\},
\label{eq:Vgeneric}
\end{equation}  
where $V_i$ is a velocity function which can be derived from the Geometric Thin-Film Equation.
Thus, the screened free-surface height admits a regular solution $\myhbar(x,t)=\sum_{i=1}^N w_i \Phi(x-x_i)$.  In this context, the weights $w_i$ and the delta-function centres $x_i$ (with $i\in \{1,2,\cdots,N\}$) can be viewed as pseudo-particles, which satisfy the first-order dynamics~\eqref{eq:Vgeneric}.
We will demonstrate in this article another advantage of this particle-method: it is a mesh-free method that automatically accumulates particles in regions where $|\partial_{xx}\myhbar|$ is large, thereby mimicking the effects of adaptive mesh refinement, with none of the computational overheads associated with that method.  A final key advantage of the particle method is positivity-preservation: as $\myhbar(x,t)=\sum_{i=1}^N w_i\Phi(x-x_i)$, with $w_i\geq 0$ and $\Phi\geq 0$, the numerical values of $\myhbar(x,t)$ are guaranteed never to be negative, hence the numerical method is manifestly positivity-preserving.


\subsection*{This work in the context of Environmental Fluid  Mechanics}

Before introducing the method, we place the method in the context of Environmental Fluid Mechanics, focusing in particular on droplet spreading on plant leaves.  In general, surfaces can be further classified as being (i) super-hydrophobic ($\thetaeq>150^\circ$), hydrophobic ($90^\circ<\thetaeq<150^\circ$), hydrophilic ($10^\circ<\thetaeq<90^\circ$), or super-hydrophilic ($\thetaeq<10^\circ$).  
Plant leaves display this wide range of contact angles.  Superhydrophobic surfaces have been frequently found in wetland plants, where the superhydrophobic surface prevents a buildup of water on the leaves, which could otherwise promote the growth of harmful micro-organisms and limit the gas exchange necessary for photosynthesis~\cite{koch2009review}.  

A particularly well-studied plant with superhydrophobic a leaf surface is the Lotus plant (\textit{Nelumbo nucifera}), with a contact angle of about $160^\circ$~\cite{koch2009review,cheng2005lotus}.  
The leaves of the Lotus plant also demonstrate a low contact-angle hysteresis, such that water droplets roll off the leaf surface when at a low tilt angle ($4^\circ$).  During rolling, contaminating particles are picked up by the water droplets, and are then removed with the  droplets as they roll off~\cite{koch2009review}.  
The leaf structure of the Lotus plant has been studied using Scanning Electron Microscopy, and reveals a hierarchical microstructure made up of micron-scale pillars (cell papillae) and a randomly covered by a smaller branch-like nanostructure~\cite{barthlott1997purity} (the wax crystal, with microstructures $\sim 100\,\mathrm{nm}$ in diameter)~\cite{saison2008replication}.    Such biological  microsturctures are the inspiration for engineered super-hydrophobic surfaces~\cite{saison2008replication}.

In contrast, plants with superhydrophilic leaf surfaces are often found in tropical regions~\cite{aryal2016variability}.  A water droplet that spreads on a superhydrophilic surface will spread and form a wide, flat droplet -- effectively a thin film.  Evaporation in such films is more efficient than in a spherical droplet, due to the increase of the water-air interface.  Thus, water evaporates from a 
superhydrophilic leaf much faster than that from a hydrophilic or superhydrophobic
one, thereby keeping the leaf dry, reducing the accumulation of harmful micro-organisms on the leaf surface, and increasing the gas exchange with the environment, for the purpose of photosynthesis~\cite{koch2009review}.  A well-studied hydrophilic plant is \textit{Ruellia devosiana}, wherein the superhydrophilic property of the leaf surface is  due both to the leaf microstructure, and to a secretion of surfactants by the leaf, which both promote spreading~\cite{koch2009superhydrophilic}.

\begin{figure}
\centering
\includegraphics[width=0.4\linewidth]{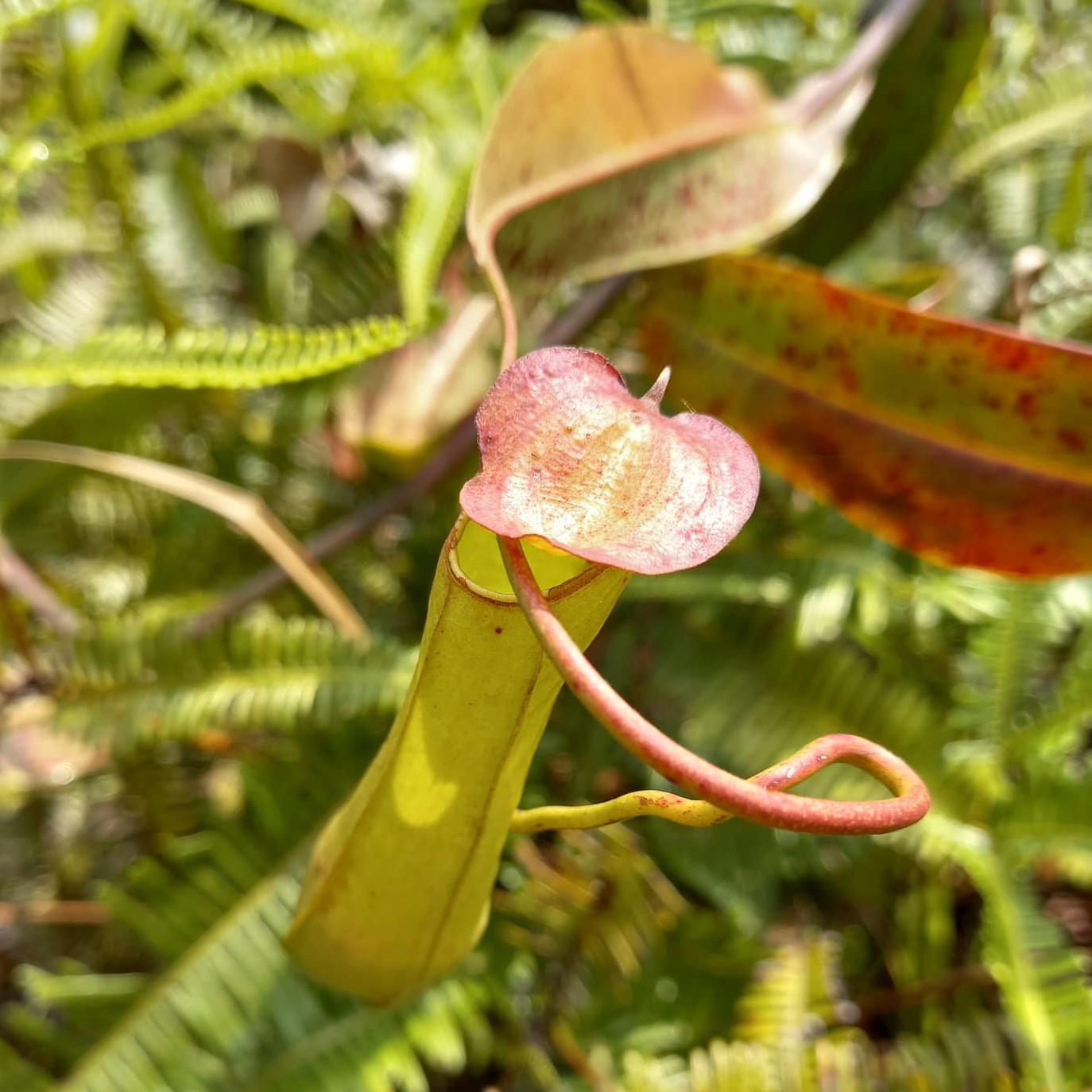}
\caption{A species of carnivorous plants of the genus Nepenthes (image: Khang Ee Pang)}
\label{fig:pitcher}
\end{figure}

Other plants which exploit hydrophilicity are the carnivorous  plants of the \textit{Nepenthes} genus (e.g. Figure~\ref{fig:pitcher}), the perisotone  of which is a fully wettable, water-lubricated anisotropic surface~\cite{bohn2004insect}.  Insects landing on the peristone or rim of the plant effectively `aquaplane' down the plant rim~\cite{bohn2004insect} before being captured by the viscoelstic fluid inside the pitcher~\cite{gaume2007viscoelastic}.

We emphasize here that both the super-hydrophobic and super-hydrophilic plant surfaces are extreme cases. 
In a comprehensive investigation of 396 plant species out of 85 families growing in three different continents at various elevations (including 792 leaf surfaces total), only forty leaf surfaces (5.1\% out of 792) were found in these extremes, including 24 super-hydrophobic surfaces and 16 super-hydrophilic surfaces~\cite{aryal2016variability}.  Thus, most leaf surfaces lie inside these extremes.
The droplet-model introduced in this paper is most relevant to hydrophilic cases where the equilibrium contact angle is small.  

\subsection*{Plan of the paper}

In Section~\ref{sec:classical} we present the classical Thin-Film Equation as a model of droplet spreading.  Certainly, this is a very well-established topic, however, we include a summary of this topic here because it enables us to clearly mark out the point of departure of the present work.  Thereafter, in Section~\ref{sec:theory} we  introduce the Geometric Thin-Film Equation as a regularized model which enables contact-line motion.  In Section~\ref{sec:particles} we introduce the particle method for generating numerical solutions of the Geometric Thin-Film Equation.   In Section~\ref{sec:complete} we presents results for complete wetting.  In Section~\ref{sec:partial} we demonstrate how the Geometric Thin-Film Equation can be extended to the case of partial wetting, and we present numerical results for that case also.  Concluding remarks are given in Section~\ref{sec:conclusions}.


\section{Review of Classical Theory}
\label{sec:classical}

In this section we review the classical theory of the Thin Film Equation, including the problem of the contact-line singularity.   The purpose of this review is to put the Geometric Thin-Film Equation into the context of the classical theory;  the theoretical formulation of the Geometric Thin-Film Equation is therefore presented subsequently in Section~\ref{sec:theory}.

\subsection*{Classical Thin-Film Equation}

We review the derivation of the classical Thin Film Equation for a flow in two dimensions, with spatial coordinates $x$ and $z$ (we refer the reader Reference~\cite{oron1997long} for the details).  The starting-point is the kinematic condition valid on the free surface $z=h(x,t)$:
\begin{equation}
\frac{\partial h}{\partial t}+u(x,z=h,t)\frac{\partial h}{\partial x}=w(x,z=h,t).
\label{eq:fs1}
\end{equation}
The fluid flow is assumed to be incompressible, such that $u_x+v_y=0$.  The incompressiblity condition can be integrated once to give
\begin{equation}
w(x,z=h,t)=-\int_0^h \frac{\partial u}{\partial x}\mathd z,\qquad z(x,z=0,t)=0.
\label{eq:incomp1}
\end{equation}
Equations~\eqref{eq:fs1}--\eqref{eq:incomp1} can be combined to give
\begin{equation}
\frac{\partial h}{\partial t}+\frac{\partial q}{\partial x}=0,\qquad q=\int_0^h u(x,z,t)\,\mathd z.
\label{eq:flux}
\end{equation}
In the lubrication limit, the velocity $u(x,z,t)$ satisfies the equations of Stokes flow, hence 
\begin{equation}
-\frac{\partial P}{\partial x}+\mu\frac{\partial^2u}{\partial z^2}=0,\qquad \frac{\partial P}{\partial z}=0,
\label{eq:stokes}
\end{equation}
 where $P$ is the fluid pressure and $\mu$ is the constant dynamic viscosity. We integrate the first equation of the pair in~\eqref{eq:stokes} once with respect to $z$ to obtain
\begin{equation}
\frac{\partial u}{\partial z}\bigg|_{z}^{h}=\frac{1}{\mu}\frac{\partial P}{\partial x}\left(h-z\right).
\label{eq:stokes1}
\end{equation}
The standard interfacial condition is that the viscous stress $\partial u/\partial z$ should vanish on the free surface $z=h(x,t)$. Thus, Equation~\eqref{eq:stokes1} becomes $\partial u/\partial z=\mu^{-1}(\partial p/\partial x)(z-h)$.  Applying the no-slip boundary condition 
\begin{equation}
u(x,z=0,t)=0, 
\label{eq:unoslip}
\end{equation}
the $u$-velocity profile becomes 
\begin{equation}
u(x,t)=\frac{1}{\mu}\frac{\partial P}{\partial x}\left(\tfrac{1}{2}z^2-h z\right),
\end{equation}
hence
\begin{equation}
q=-\frac{1}{3\mu}h^3\frac{\partial P}{\partial x}.
\label{eq:qdef}
\end{equation}
The pressure $P$ is identified with the Laplace pressure, $P=-\surften \partial_{xx}h$, where $\surften$ is the surface tension and $h_{xx}$ is the interfacial curvature in the longwave limit.  Hence, Eqution~\eqref{eq:qdef} becomes:
\begin{equation}
q=\frac{\surften}{3\mu}h^3h_{xxx}.
\label{eq:qdef1}
\end{equation}
Substituting Equation~\eqref{eq:qdef1} into Equation~\eqref{eq:flux} gives:
\begin{equation}
\frac{\partial h}{\partial t}+\frac{\surften}{3\mu}\frac{\partial}{\partial x}\left(h^3\frac{\partial^3 h}{\partial x^3}\right)=0.
\label{eq:hdef}
\end{equation}

\subsection*{Contact-Line Singularity}

In the context of droplet spreading, it is desirable to propose a similarity solution to Equation~\eqref{eq:hdef}, corresponding to a self-similar droplet that retains some overall structural properties even as the base of the droplet spreads out.  Dimensional analysis indicates that the similarity solution should be:
\begin{equation}
h(x,t)=h_0\left(t/t_0\right)^{-1/7} f(\eta),\qquad \eta=\frac{x/R}{(t/t_0)^{1/7}},
\label{eq:similarity1}
\end{equation}
where $h_0$ and $R$ are as given in Figure~\ref{fig:sketch1} and $t_0$ is a timescale to be determined.  Substitution of Equation~\eqref{eq:similarity1} into Equation~\eqref{eq:hdef} yields:
\begin{equation}
\tfrac{1}{7}\eta f=\frac{\surften}{3\mu}\frac{t_0}{h_0}\left(\frac{h_0}{R}\right)^3f^3f'''.
\label{eq:similarity2}
\end{equation}
The timescale $t_0$ is chosen to be the capillary timescale, such that $(\surften/3\mu)(t_0/h_0)(h_0/R)^3=1$.  Thus, Equation~\eqref{eq:similarity2} becomes:
\begin{equation}
\tfrac{1}{7}\eta f=f^3f'''.
\label{eq:similarity3}
\end{equation}
The appropriate droplet-spreading boundary conditions for Equation~\eqref{eq:similarity3} are $f(0)=1$, $f'(0)=0$, 
\begin{equation}
f=f'=0,\qquad \text{at }\eta=\eta_0>0,
\label{eq:similarity_bc}
\end{equation}
 where $\eta_0$ corresponds to the outermost extent of the droplet.  Thus, the position $a$ at which the (microscopic) contact line touches  down to zero is described by $a(t)/R=\eta_0 (t/t_0)^{1/7}$.   Unfortunately, the similarity solution~\eqref{eq:similarity1} with the boundary conditions~\eqref{eq:similarity_bc} fails to exist; instead, $f(\eta)$  degenerates into a Dirac delta function centred at $\eta=0$, and the droplet does not spread~\cite{hulshof2001some}.  

The reason for this failure is that the no-slip condition~\eqref{eq:unoslip} is inconsistent with the phenomenon of droplet spreading.  When the model~\eqref{eq:hdef} is applied to droplet spreading, the physics which permits  slip to occur on sufficiently small scales is missing.  
The missing physics is then put into the model as part of a regularization.  For instance, by allowing for slip on a sufficiently small scale $\ell$, Equation~\eqref{eq:hdef} becomes:
\begin{equation}
\frac{\partial h}{\partial t}+\frac{\surften}{\mu}\frac{\partial}{\partial x}\left[\left(\tfrac{1}{3}h^3+\ell h^2\right)\frac{\partial^3 h}{\partial x^3}\right]=0.
\label{eq:hdefslip}
\end{equation}
Equation~\eqref{eq:hdefslip} is the Thin-Film Equation with a \textit{slip-length model}.  

Using the theory of matched asymptotic expansions, it has been shown~\cite{hocking1982technical} that the solution of Equation~\eqref{eq:hdefslip} consists of an outer solution and an inner solution.  The outer solution resembles the similarity solution~\eqref{eq:similarity1} and is valid on large scales, far from the contact line.  The inner solution is valid on small scales close to the contact line, and provides for contact-line motion.  Indeed, by matching the inner and outer solutions across an intermediate matching zone, the contact-line $a(t)$ is shown to satisfy the so-called \textit{Tanner's Law},
\begin{equation}
\frac{\mathd a}{\mathd t}=\frac{\surften\theta_0^3R^6}{3\mu}\epsilon\left[1+\epsilon\left(2-\ln\frac{2a}{R}\right)\right]a^{-6}.
\end{equation} 
Where $\theta_0$ is the initial contact angle. Thus, $a(t)\sim t^{1/7}$, which is the scaling that would be expected if the similarity solution could be made to extend down to the microscopic scale.

The slip-length model therefore provides for a resolution of the contact-line singularity.  However, the stress $\surften h_{xx}$ remains singular at the contact line.  For these reasons, an alternative regularization of the Thin-Film Equation has been proposed, namely the Precursor-Film model~\cite{de1985wetting,bonn2009wetting}.    Following Reference~\cite{holm2020gdim}, in this work we present the Geometric Thin-Film Equation as an alternative regularization, the advantage of this approach as we reveal in subsequent sections is the remarkable simplicity of the numerical solutions produced by this model.

\section{Geometric Thin-Film Equation: Theoretical Formulation}
\label{sec:theory}

In the framework of the Geometric Thin-Film Equation, the starting-point is the assumption that there is missing physics on a small scale.  Instead of modelling the missing physics, it is parametrized.  As such, $h(x,t)$ is used to denote the interface location in a crude model with missing physics -- which we call here the `noisy' interface location.  The noisy interface location is to be smoothened by a filtering operation, to produce a smoothened, more accurate, estimate of the interface location, which we denote by $\myhbar(x,t)$.  The noisy interface location may be different from the true interface location -- for instance, the noisy interface location may be zero, whereas the true interface location may be close to, but different from zero -- as would be the case if the noisy interface location was obtained through an incomplete model with missing small-scale physics.

\subsection*{Model A}

To take account of the fact that $\myhbar$ represents a smoother description of the interface location than $h$, we propose that $h$ and $\myhbar$ be connected via the expression
\begin{equation}
h=\myhbar+\eta,
\label{eq:uncertainty}
\end{equation}
where $\eta$ is a fluctuating quantity with mean zero and variance $\sigma^2$.  Then, $\myhbar(x,t)$ can be made into an accurate estimate of the interface location by minimizing the total interfacial energy
\begin{equation}
\Energy[\myhbar]=\gamma\int_{-\infty}^\infty \sqrt{1+|\partial_x\myhbar|^2}\mathd x,
\label{eq:Adef}
\end{equation}
subject to a fixed-variance-constraint:
\begin{equation}
\int_{-\infty}^\infty |h-\myhbar|^2\mathd x=\sigma^2.
\label{eq:constop}%
\end{equation}%
Here, $\gamma$ is a positive constant representing the surface area; the constraint~\eqref{eq:constop} enforces a fixed level of uncertainty between the model with missing physics and the smoothened model.
In practice, we minimize the surface area in the long-wave limit: instead of Equation~\eqref{eq:Adef} we minimize
\begin{equation}
\Energy[\myhbar]=\tfrac{1}{2}\surften\int_{-\infty}^\infty |\partial_x\myhbar|^2\mathd x,
\label{eq:Along}
\end{equation}
which is obtained from Equation~\eqref{eq:Adef} in the longwave limit, when $|\partial_x\myhbar|^2$ is small.

Equation~\eqref{eq:Along} with constraint~\eqref{eq:constop} is a constrained minimization problem -- to solve it, one would introduce an energy functional with a Lagrange multiplier:
\begin{equation}
\Lagrange[h,\myhbar]=\Energy[\myhbar]+\lambda \left[\tfrac{1}{2}\int_{-\infty}^\infty |h-\myhbar|^2\mathd x-\tfrac{1}{2}\sigma^2\right].
\label{eq:multiplier}
\end{equation}
One would then compute
\begin{equation}
\frac{\delta \Lagrange}{\delta \myhbar}=0,\qquad \frac{\delta \Lagrange}{\delta h}=0,
\end{equation}
yielding
\begin{equation}
-\gamma\partial_{xx}\myhbar-\lambda (h-\myhbar)=0,\qquad h-\myhbar=0.
\label{eq:lagrange}
\end{equation}
In practice, solving Equation~\eqref{eq:lagrange} yields inconsistent results, as it implies that $h=\myhbar$.  But $h$ and $\myhbar$ live in different function spaces ($h$ is noisy, $\myhbar$ is smooth), so Equation~\eqref{eq:lagrange} cannot be correct.  
Instead, we can study the dynamics, whereby $\Lagrange[h,\myhbar]$ in Equation~\eqref{eq:multiplier} gradually evolves to a minimum configuration over time.  The dynamics are highly conditioned:
\begin{itemize}
\item The evolution of $h$ and $\myhbar$ must be such that $\Lagrange$ tends to a minimum over time;
\item The integrals $\int_{-\infty}^\infty h(x,t)\mathd x$ and $\int_{-\infty}^\infty \myhbar(x,t)\mathd x$ must be conserved quantities, reflecting underlying principles of conservation of fluid mass.
\end{itemize}
Under these conditions, the evolution equation for $h$ must be of a generic conservative-gradient-descent type, hence:
\begin{equation}
\frac{\partial h}{\partial t}=\frac{\partial}{\partial x}\left(hM\frac{\partial}{\partial x}\frac{\delta \Lagrange}{\delta h}\right).
\label{eq:evolution_a}
\end{equation}
where $M\geq 0$ is a mobility function to be determined.  The evolution equation for $\myhbar$ may be similar.  However, for simplicity, we may assume that $\myhbar$ relaxes instantaneously to a smoothened form of $h$, hence $\delta \Lagrange/\delta\myhbar=0$, hence
\begin{equation}
-\gamma\partial_{xx}\myhbar=\lambda(h-\myhbar),
\label{eq:helmholtz1}
\end{equation}
or
\begin{equation}
\myhbar=\left(1-\frac{\gamma}{\lambda}\partial_{xx}\right)^{-1}h.
\label{eq:helmholtz2}
\end{equation}
Equation~\eqref{eq:helmholtz2} establishes a natural smoothing operation and hence, smoothing kernel for the formulation, namely, the Helmholtz kernel.
Substitution of Equation~\eqref{eq:helmholtz1} into Equation~\eqref{eq:evolution_a} yields:
\begin{equation}
\frac{\partial h}{\partial t}=-\frac{\partial}{\partial x}\left[hM\frac{\partial}{\partial x}\left(\surften\partial_{xx}\myhbar\right)\right].
\label{eq:evolution_b}
\end{equation}
The physical model for $h$ is completed by specifying the mobility.  This is done by reference to the classical theory in Section~\ref{sec:classical}.  However, instead of $M=(1/3\mu)h^2$ we take
\begin{equation}
M=\frac{1}{3\mu}\myhbar^2;
\label{eq:mobility}
\end{equation}
the reason for using $\myhbar^2$ in the mobility becomes apparent when we look at particle-like solutions of the regularized model (Section~\ref{sec:particles}).
Finally, the value of the Lagrange multiplier $\lambda$ is chosen at each point in time to reflect the model uncertainty:
\begin{equation}
\int_{-\infty}^\infty \left|\left(1-\frac{1}{\lambda}\partial_{xx}\right)^{-1}h-h\right|^2\mathd x=\sigma^2.
\label{eq:lagrange_solve}
\end{equation}
We refer to this model with a fixed level of uncertainty as Model A.

\subsection*{Model B}

In practice, recomputing the Lagrange multiplier $\lambda$ at each time $t$ is a difficult task numerically.  However, an equivalent model can be formulated 
by introducing an unconstrained functional,
\begin{equation}
\Lagrange[h,\myhbar]=\Energy[\myhbar]+\tfrac{1}{2}\frac{\gamma}{\alpha^2}\int_{-\infty}^\infty (h-\myhbar)^2\mathd x.
\end{equation}
Here, the parameter $\alpha$ corresponds to model uncertainty on the (small) lengthscale $\lambda$.  The dynamical equation is the same as before (Equation~\eqref{eq:evolution_b}), as is the mobility; however, now $\myhbar$ is computed as 
\begin{equation}
\myhbar=\left(1-\alpha^2\partial_{xx}\right)^{-1}h:=K*h.
\label{eq:helmholtz3}
\end{equation}
We refer to this model with uncertainty on a small scale $\alpha$ as Model B.  Here,  we have introduced the standard notation for smoothing kernels:
\[
K*f(x)=(1-\alpha^2\partial_{xx})^{-1}f(x)=\int_{-\infty}^\infty K(x-y)f(y)\mathd y,
\]
and we explicitly use $K$ for the Helmholtz kernel, such that
\[
K(x)=\frac{1}{2\alpha}\mathe^{-\alpha|x|}.
\]

Although Model A and Model B are different, there is a one-to-one relationship between them, and they are equivalent -- e.g. $\lambda$ in Equation~\eqref{eq:lagrange_solve} is clearly a $\left[\text{lengthscale}\right]^2$ which depends on time.  We therefore identify
\begin{equation}
\alpha(t)=\left[\lambda(t;\sigma)\right]^{-1/2},
\label{eq:lagrange_solve1}
\end{equation}
and the required uncertainty on a small lengthscale $\alpha$ in the second description of the model is the average value of Equation~\eqref{eq:lagrange_solve1}:
\begin{equation}
\alpha=\lim_{T\rightarrow\infty}\frac{1}{T}\int_0^{\infty}\left[\lambda(t;\sigma)\right]^{-1/2}\mathd t.
\end{equation}
Due to the computational efficiency, Model B is preferred in this work.

The kernel solution~\eqref{eq:helmholtz3} can be substituted back into the expression $\Lagrange[h,\myhbar]$ to give:
\begin{equation}
\lagrange[h]:=\Lagrange\left[h,\myhbar=\left(1-\alpha^2\partial_{xx}\right)^{-1}h\right]=\tfrac{1}{2}\gamma\int_{-\infty}^\infty \left[\left(\partial_x\myhbar\right)^2+\alpha^2\left(\partial_{xx}\myhbar\right)^2\right]\mathd x.
\label{eq:ell1}
\end{equation}
This can in turn be written in several further ways:
\begin{enumerate}
\item The inner-product pairing of $\partial_x h$ with $\partial_x \myhbar$:
\begin{equation}
\lagrange[h]=\tfrac{1}{2}\gamma\int_{-\infty}^\infty \partial_x h \partial_x\myhbar\,\mathd x.
\label{eq:ell2}
\end{equation}
\item The weighted inner-product pairing:
\begin{equation}
\lagrange[h]=\tfrac{1}{2}\gamma\langle \partial_x\myhbar,\partial_x\myhbar\rangle_K=\tfrac{1}{2}\gamma\int_{-\infty}^\infty \partial_x \myhbar\left(1-\alpha^2\partial_x^2\right) \partial_x\myhbar\,\mathd x.
\label{eq:ell3}
\end{equation}
The pairing $\langle \cdot,\cdot\rangle_K$ defines a Reproducing Kernel Hilbert Space~\cite{evgeniou2000regularization}.
\end{enumerate}
Equation~\eqref{eq:evolution_b} now reads:
\begin{equation}
\frac{\partial h}{\partial t}=-\frac{\partial}{\partial x}\left(hM\frac{\partial}{\partial x}\frac{\delta \lagrange}{\delta h}\right),\qquad \myhbar=K*h.
\label{eq:evolution_c}
\end{equation}

\subsection*{Higher-order smoothing}

For the purpose of generating particle-like solutions of the regularized model (e.g. Section~\ref{sec:particles}), smoothing with the Helmholtz kernel is not sufficient.  Therefore, in this paper, we work with a higher-order smoothing.  We take $\ell$ as before (specifically, Equation~\eqref{eq:ell1}), with evolution equation~\eqref{eq:evolution_c} and smoothing kernel $\myhbar=K*K*h$ -- this is a straightforward extension of the basic model.  We therefore summarize in one place the model studied in this work:
\begin{subequations}
\begin{eqnarray}
\lagrange[h]&=&\tfrac{1}{2}\gamma \int_{-\infty}^\infty \partial_x h\partial_x\myhbar\,\mathd x,\\
\myhbar&=&K*K*h,\\
\frac{\partial h}{\partial t}&=&-\frac{\partial}{\partial x}\left[hM\frac{\partial}{\partial x}\left(\partial_{xx}\myhbar\right)\right].
\end{eqnarray}%
\label{eq:particle}%
\end{subequations}%
The aim of the remainder of the paper is to explore numerically the solutions of Equation~\eqref{eq:particle} -- we will use $\Phi=K*K$ to denote the double Helmholtz kernel, such that $\myhbar=\Phi*h$ -- the need for this higher-order smoothing will become apparent in Section~\ref{sec:particles}.

\subsection*{Discussion}

Summarizing our work so far, we have introduced a regularized thin-film equation where the missing small-scale physics is not modelled, but is instead parametrized. The idea of the model is that $h(x,t)$ provides an incomplete description of the droplet evolution (but which nonetheless contains important physical information, such as the problem dependence on the viscosity $\mu$ and time $t$.  A refined description of the interface is then obtained via a smoothened interface profile $\myhbar=\Phi*h$.  Overall, the model evolves so as to minimize the interfacial energy (area) while keeping the difference between $h$ and $\myhbar$ as small as possible.

The model as formulated envisages that $h(x,t)$ is an incomplete description of the interface profile (with missing small-scale physics).  The missing physics is encoded either as a fixed level of uncertainty between $h$ and $\myhbar$ (model A), or such that the uncertainty in the model description occurs below a lengthscale $\alpha$ (model B).  These two models are equivalent, although model B is preferred for computational simplicity.

The equation~\eqref{eq:particle} is a variant of the so-called Geometric Thin-Film equation introduced in Reference~\cite{holm2020gdim}: 
by viewing $a=h\mathd x$ as a one-form, Equation~\eqref{eq:particle} can be written as
\begin{equation}
\frac{\partial a}{\partial t}=-\lie_{U}(a),
\label{eq:lie3}
\end{equation}
where $\lie_U(a)=\lie_U(h\mathd x)=\partial_x(U h)\,\mathd x$ is the Lie derivative on one-forms; in this instance, $U=M\partial_x(\delta \lagrange/\delta h)$ is the pertinent vector field.  Equation~\eqref{eq:lie3} is then a very simple instance of a general theory of gradient-flow dynamics~\cite{holm2008formation} which uses geometric mechanics (Lie Derivatives, Momentum Maps) to formulate an energy-dissipation mechanical model for general configuration spaces -- hence the \textit{Geometric} Thin-Film equation.
Equation~\eqref{eq:particle}--\eqref{eq:lie3} can furthermore be identified as a special case of Darcy's Law~\cite{holm2008formation}, where the generalized force is $f=-\partial_x(\delta \lagrange/\delta h)$, the Darcy velocity is $U=Mf$, and the flux-conservative evolution is equation for the conserved scalar quantity $h$ is $h_t+\partial_x(hU)=0$.  These insights will be used in formulating the Geometric Thin-Film equation for the case of partial wetting in Section~\ref{sec:partial}, below.

We remark finally here on the intriguing connection between Model A and `denoising' in Image Processing -- in Image Processing one is given a noisy image $h$ and it is desired to produce a smoother image $\myhbar$ while keeping the difference between the noisy and the smooth image at a fixed level $\sigma$.  This is achieved by minimizing a functional such as Equation~\eqref{eq:multiplier}~\cite{karkkainen2005denoising,diffellahimage}.  Our evolution equation~\eqref{eq:evolution_b} for the free-surface height $h(x,t)$ in a thin-film flow is equivalent to carrying out denoising on a (one-dimensional) image in Image Processing.

\section{Geometric Thin-Film Equation: Particle Solutions}
\label{sec:particles}

The general theory of geometric dissipative mechanics introduced in Reference~\cite{holm2008formation} includes many examples where discrete, particle-like solutions (such as Equation~\eqref{eq:Vgeneric}) are admitted.  Motivated by these examples, we seek simplified solutions of Equation~\eqref{eq:particle} of the following form:
\begin{subequations}
\begin{eqnarray}
h^N&=&\sum_{i=1}^N w_i \delta(x-x_i(t)),\\
\myhbar^N(x,t)&=&\sum_{i=1}^N w_i \Phi(x-x_i(t)).\label{eq:hNdefb}
\end{eqnarray}%
\label{eq:hNdef}%
\end{subequations}%
where $N$ is a positive integer corresponding to a truncation of an infinite sum, $w_i\geq 0$ are weights to be computed, and $\delta(\cdot)$ is the Dirac delta function.  The motivation for seeking out such highly simplified particle solutions is that they make the task of solving the partial differential equation~\eqref{eq:particle} numerically very simple: instead of discretizing a fourth-order parabolic-type partial differential equation and solving it numerically, we can instead solve a set of ordinary differential equations for the delta-function centres $x_i(t)$ using standard time-marching algorithms.   This simplifies the numerical computations greatly.    We refer to the weights $w_i$ together with the delta-function centres $x_i(t)$ as the `particles' -- thus, we are concerned with a particle-solution of Equation~\eqref{eq:particle}.  We describe the construction of such particle-solutions in what follows.

The weights $w_i\geq 0$ are chosen such that
\begin{equation}
\lim_{N\rightarrow\infty} \int_{-\infty}^\infty h^N(x,t=0)\phi(x)\mathd x=\int_{-\infty}^\infty h_0(x)\phi(x)\mathd x,
\label{eq:compatibity1}
\end{equation}
where $\phi(x)$ is an arbitrary smooth, integrable test function.  This limit can be satisfied by taking
\begin{equation}
x_i(t=0)=x_i^0=\left(i-\frac{N}{2}\right)\frac{2L}{N},\qquad i\in\{1,2,\cdots\,N\},
\label{eq:compatibity_x0}
\end{equation}
where $L$ is a lengthscale such that $\mathrm{supp}(h_0)\subset [-L,L]$, and by taking
\begin{equation}
w_i=h_0(x_i^0)(2L/N),\qquad i\in\{1,2,\cdots\,N\},
\label{eq:compatibity_wi}
\end{equation}
then, Equation~\eqref{eq:compatibity1} is satisfied automatically.

 We now multiply both sides of Equation~\eqref{eq:particle} by the test function $\phi(x)$, integrate from $x=-\infty$ to $x=\infty$, and apply vanishing boundary conditions at these limits.  We thereby obtain
\begin{equation}
\langle \phi ,h_t\rangle-\frac{\surften}{3\mu}\langle \phi_x,h \myhbar^2 \partial_{xxx}\myhbar\rangle=0,
\label{eq:weak1}
\end{equation}
where $\langle \cdot,\cdot\rangle$ denotes the standard pairing of square-integrable functions:
\begin{equation}
\langle f,g\rangle=\int_{-\infty}^\infty f g \mathd x,\qquad f,g, \in L^2(\mathbb{R}).
\end{equation}
We substitute Equations~\eqref{eq:hNdef} into Equation~\eqref{eq:weak1}.  Owing to the judicious choice of mobility $\overline{M}=(1/3\mu)\myhbar^2$ (\textit{cf.} Equation~\eqref{eq:mobility}), no instance of the singular solution $h^N(x,t)$ gets squared (only the smoothened solution $\myhbar^N$ gets squared).  After performing some standard manipulations with Dirac delta functions, Equation~\eqref{eq:weak1} becomes:
\begin{equation}
\sum_{i=1}^N w_i\phi_x(x_i)\frac{\mathd x_i}{\mathd t}-\frac{\surften}{3\mu}\sum_{i=1}^N w_i \phi_x(x_i) \left[(\myhbar^N)^2 \partial_{xxx}\myhbar^N\right]_{x=x_i}=0,
\label{eq:weak2}
\end{equation}
where now $[(\myhbar^N)^2 \partial_{xxx}\myhbar^N]_{x=x_i}$ is taken to mean
\begin{equation}
\bigg\{\left[\sum_{j=1}^N w_j \Phi(x-x_j)\right]^2\left[\sum_{j=1}^N w_j \Phi'''(x-x_j)\right]\bigg\}_{x=x_i}.
\end{equation}
Equation~\eqref{eq:weak2} is re-arranged to give
\begin{equation}
\sum_{i=1}^N w_i\phi_x(x_i)\left[ \frac{\mathd x_i}{\mathd t}-\frac{\surften}{3\mu}\left[(\myhbar^N)^2 \partial_{xxx}\myhbar^N\right]_{x=x_i}\right]=0.
\label{eq:weak3}
\end{equation}
Equation~\eqref{eq:weak3} is true for all test functions $\phi(x)$ and all initial data $h_0(x)$ (hence $w_i$), hence 
\begin{equation}
\frac{\mathd x_i}{\mathd t}-\frac{\surften}{3\mu}\left(\myhbar^2 \partial_{xxx}\myhbar\right)_{x=x_i}=0.
\label{eq:ode}
\end{equation}
Thus, Equation~\eqref{eq:hNdef}, together with the ordinary differential equations
\begin{equation}
\frac{\mathd x_i}{\mathd t}=\frac{\surften}{3\mu}\left(\myhbar^2 \partial_{xxx}\myhbar\right)_{x=x_i},\qquad t>0,\qquad i=1,2,\cdots,N,
\label{eq:odedefx}
\end{equation}
and initial data
\begin{equation}
x_i(t=0)=x_i^0 =\left(i-\frac{N}{2}\right)(2L/N),\qquad \mathrm{supp}(h_0)\subset [-L,L],
\end{equation}
give a  so-called \textit{singular solution} to the Geometric Thin-Film Equation~\eqref{eq:particle}.  The centres of the Dirac delta functions $x_i(t)$ with associated weights $w_i$ are identified as pseudo-particles, and the velocity $V_i$ of the $i^{\text{th}}$ pseudo-particle is identified with the right-hand side in  Equation~\eqref{eq:odedefx},
\begin{equation}
\frac{\mathd x_i}{\mathd t}=V_i(x_1,\cdots,x_N),\qquad 
V_i(x_1,\cdots,x_N)=\frac{\surften}{3\mu}\left(\myhbar^2 \partial_{xxx}\myhbar\right)_{x=x_i}.
\label{eq:odedefx1}
\end{equation}

\subsection*{Key properties of the particle evolution equations}

We notice that in Equation~\eqref{eq:odedefx1}, evaluation of $\Phi'''$ is required -- this is the rationale for our choice of the double Helmholtz kernel as the smoothing kernel in 
Equation~\eqref{eq:particle}.  Using the single Helmholtz kernel $K$ would not be sufficient, as $K'''$ is singular at the origin.  Furthermore, as the reconstructed interface profile $\myhbar(x,t)=\sum_{i=1}^Nw_i\Phi(x-x_i(t))$ involves the positive weights $w_i$ and a positive kernel $\Phi\geq 0$, the particle method is positivity-preserving: if $h$ and $\myhbar$ are initially positive, then the stay positive for all time.  The numerical particle method is manifestly positivity-preserving, this is a key advantage as a numerical method that led to erroneous negative values of $h$ and $\myhbar$ would produce unphysical results.

\subsection*{Numerical Solutions using the Particle Method}

In this paper, we solve the Geometric Thin-Film Equation~\eqref{eq:particle} 
numerically using the particle method~\eqref{eq:odedefx1}.   To demonstrate the accuracy of the novel particle method, we compare the particle method to a standard fully-implicit finite-difference method.  It can be noted from the particle method (specifically Equation~\eqref{eq:odedefx1}) that $N$ ordinary differential equations are to be solved; each of the $N$ right-hand-side (RHS) terms requires a summation over all other particles (\textit{cf.} Equation~\eqref{eq:hNdefb} and the second entry in Equation~\eqref{eq:odedefx1}) -- this suggests the particle method has computational complexity $O(N^2)$.  However, the number of floating-point operations to be performed in evaluating the different RHS terms can be dramatically reduced by symmetry operations, to give an overall computational complexity  $O(N)$ -- this is the so-called fast-particle method.  We give details of the fast particle method and our fully-implicit finite-difference method in Appendix~\ref{sec:app:numerics}. 

Both the particle method and the fully-implicit finite-difference method are solved in Matlab.  The particle method makes use of Matlab's built-in time-evolution algorithms for ordinary differential equations, notably, ODE45 and ODE15s.


\section{Complete Wetting}
\label{sec:complete}

In this section we present numerical results for complete wetting for the Geometric Thin-Film Equation.  Equation~\eqref{eq:particle} clearly corresponds to the case of complete wetting: the energy functional $\Energy=(1/2)\gamma\int_{-\infty}^\infty |\partial_x\myhbar|^2\mathd x$ is penalized, meaning that the system evolves to minimize the curved part of the droplet interface, that is, the part of the droplet interface in contact with the surrounding atmosphere. 


\subsection*{Non-dimensionalization and Initial Conditions}

To characterize this spreading phenomenon, we solve the Geometric Thin-Film equation~\eqref{eq:particle} in dimensionless variables.  The interfacial height is made dimensionless on a lengthscale $h_0$, which is proportional to the initial maximum droplet height $h_{max}$.
In this section, we take the rather non-standard value $h_0=(8/3)h_{max}$ for $h_0$, this is done here to compare with Reference~\cite{holm2020gdim}.
The $x$-coordinate is then made dimensionless on the initial droplet base $L$.  
Finally, time is made dimensionless on the capillary timescale $\tau=(3\mu L/\gamma)(L/h_0)^3$.  The ratio $\epsilon=h_0/L$ must be small, for inertial effects to be negligible, and hence, for the lubrication theory underlying Equation~\eqref{eq:particle} to be valid.
Henceforth, we assume that all of the relevant variables have been made dimensionless in this way.  The initial condition for the droplet height therefore reads:
\begin{align}
    h(x,t=0)=
    \begin{cases}
        \frac{3}{2}\left[\left(\tfrac{1}{2}\right)^2-x^2\right], &\qquad \text{if }|x|<\tfrac{1}{2}, \\
        0, &\qquad \text{otherwise}.
    \end{cases}
		\label{eq:init1}
\end{align}
Thus, the droplet area (which is the analogue of droplet volume in two dimensions) is therefore fixed as
\[
\int h(x,t=0)\,\mathd x=1/4.
\]
Both the droplet area $\int h(x,t)\,\mathd x$ and $\int \myhbar(x,t)\,\mathd x$ are conserved under the evolution equation~\eqref{eq:particle}.  The simulations are carried out in a finite spatial domain $x\in [-2,2]$ with periodic boundary conditions -- this condition also establishes the limits of integration on the foregoing integrals.

\subsection*{Results}

We solve Equation~\eqref{eq:particle} with the initial condition~\eqref{eq:init1}.
The numerical calculations indicate that the particle method and the finite-difference method produce results that are qualitatively the same: we therefore show only results for the particle method.  A rigorous, quantitative comparison between the two methods is also presented herein -- this analysis also justifies the number of particles used in the calculations as $N=800$, this can be deemed equivalent to a grid spacing of $\Delta x=2L/N=0.005$ in the finite-difference method.

\begin{figure}[h]
	\begin{minipage}{0.45\textwidth}
        \centering
        \includegraphics[width=\linewidth]{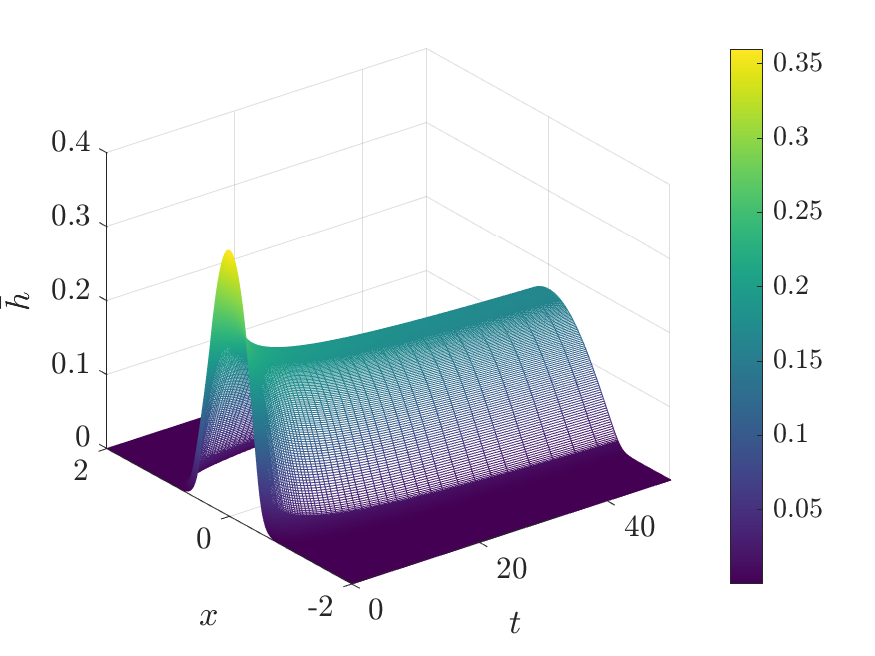}
    \end{minipage}
    \begin{minipage}{0.45\textwidth}
        \centering
        \includegraphics[width=\linewidth]{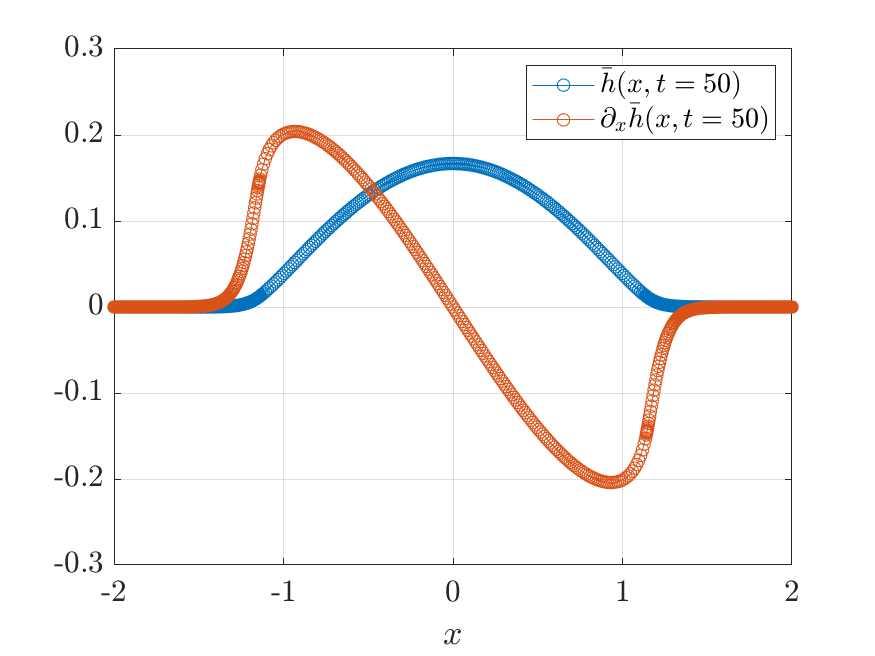}
    \end{minipage}
    \caption{(a) Space-time plot of $\bar{h}(x,t)$ showing the spreading of the droplet. (b) Droplet shape $\bar{h}$ and the slope $\partial_x\bar{h}$ at $t=50$. }
    \label{fig:solution}
\end{figure}

A first set of results is shown in Figure~\ref{fig:solution}. 
In Figure~\ref{fig:solution}(a) we show a space-time plot of the smoothened free-surface height $\bar{h}$ where the spatial grid is evaluated at the particle positions $x_i(t)$ -- effectively, a discretization of $\myhbar$ on a non-uniform grid.  From this plot, region in space where
where $\myhbar$ is significantly different from zero increases over time, demonstrating that the droplet is indeed spreading.
Figure \ref{fig:solution}(b) shows a snapshot the filtered surface height and its slope at $t=50$.  
The particle locations are shown explicitly in Figure~\ref{fig:solution}(b) -- there is a high concentration of particles  in the regions of high curvature -- this is discussed in more detail in what follows.

\begin{figure} [htb]
	\includegraphics[width=0.6\textwidth]{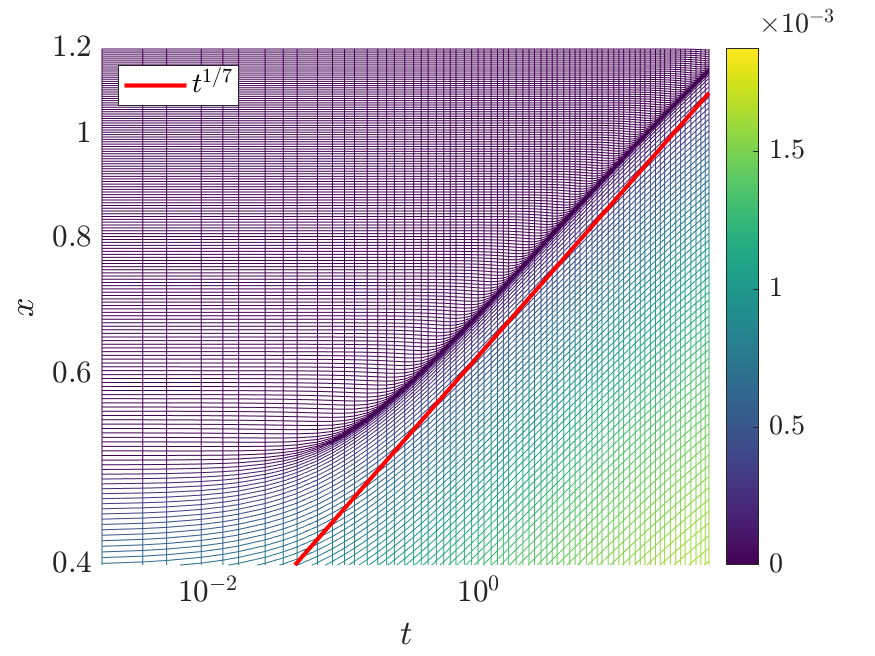}
\caption{Evolution of the particle trajectories $x_i(t)$ (logarithmic scales on both axes). The colors indicate the weight corresponding to each particle $w_i$. The line $t^{1/7}$ is imposed to show that the trajectories follows a power law at late time.}
	\label{fig:trajectory}
\end{figure}

In Figure \ref{fig:trajectory}, we plot the particle trajectories $x_i(t)$ for the solution of the Geometric Thin-Film Equation via the particle method.   
Intriguingly, the particles are seen to accumulate at regions where $|\partial_{xx}\myhbar|$  is large, giving a higher spatial resolution in regions of high interfacial curvature.  When a finite-difference or finite-volume solver is able to execute local grid refinement in regions of where the spatial derivatives are large in magnitude, this is because of adaptive mesh-refinement, which is a complex and computationally expensive feature to add to a numerical solver.  Here, the particle method demonstrates a built-in tendency to mimic the effect of adaptive mesh-refinement, without the high computational overhead of that method. 
Furthermore, in Figure~\ref{fig:trajectory} we have used the built-in MATLAB ODE solvers to generate the space-time plot, meaning that the adaptive mesh refinement is performed in the temporal as well as the spatial domain.

A further advantage of the particle method is that it provides a numerical description of the contact line by simply following the trajectory of a particle starting near the contact line, say $x_k(t)$ such that $x_k(0)=0.5$.  This is also shown in Figure~\ref{fig:trajectory}, where the contact line is found to satisfy Tanner's Law, with $x_k(t)\sim t^{1/7}$.

%

\begin{figure}[h]
        \centering
        \includegraphics[width=0.6\textwidth]{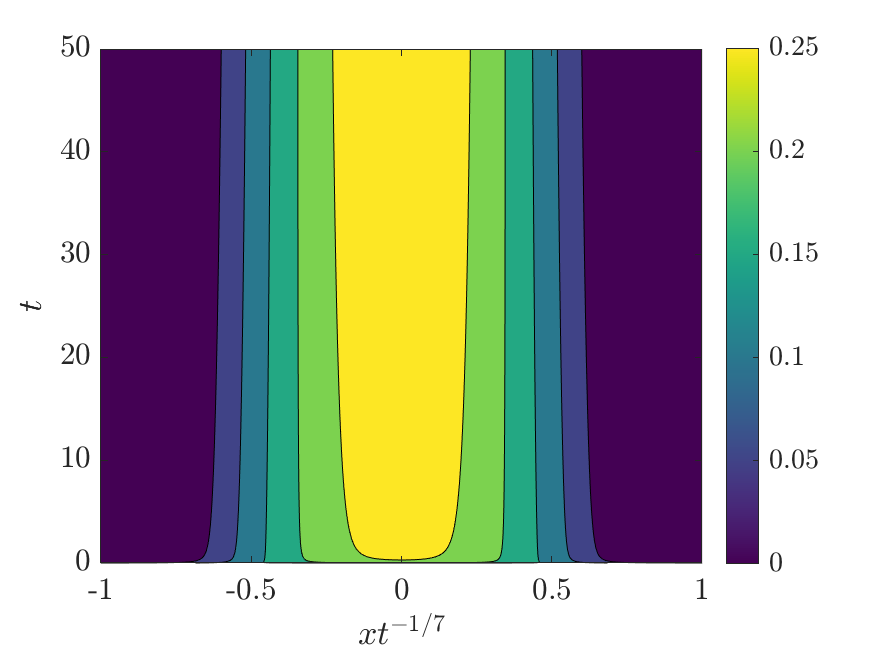}
\caption{Space-time plot in similarity variables of $f_\alpha(\eta,t)=t^{-1/7}\myhbar(\eta t^{1/7},t)$.}
\label{fig:similarity1}
\end{figure}

\begin{figure}[h]
	\centering
  \includegraphics[width=0.6\textwidth]{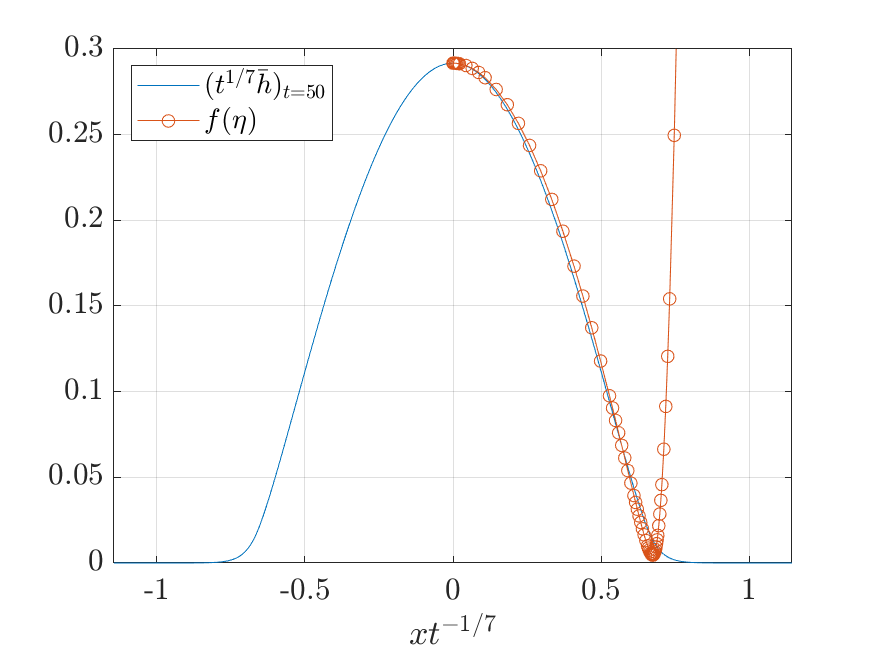}
  \caption{Comparison between $f_\alpha(\eta,t=50)$ and the similarity solution, solved using the shooting method.}
  \label{fig:similarity2}
\end{figure}

The finding that the Geometric Thin-Film Equation satisfies Tanner's Law of droplet spreading indicates that the regularized model is capturing the large-scale physics in the droplet-spreading problem.  In order to demonstrate this even further, we introduce the function
$f_\alpha(\eta,t)=t^{-1/7}\myhbar(\eta t^{1/7},t)$, with $\eta t^{1/7}=x$.  In Figure \ref{fig:similarity1} we produce a space-time plot of $f_\alpha(\eta,t)$ -- this is seen to relax to a constant profile at late times as $t\rightarrow \infty$.  
Furthermore, the  profile of $f_\alpha(\eta,t)$ at fixed $t$ ($t$ large) can be compared with a similarity solution of the unregularized problem, $f^3f'''=\eta f/7$ 
(\textit{cf.} Equation~\eqref{eq:similarity3}).  This ordinary differential equation is  then solved with the shooting method together with appropriate initial conditions~\cite{holm2020gdim}.  The results are shown in Figure~\ref{fig:similarity2}.
This figure therefore shows that the Geometric Thin-Film Equation describes the expected large-scale droplet-spreading physics in the droplet core.  Where the classical Thin-Film Equation breaks down at the contact line, the height profile of the Geometric version decays smoothly to zero. 

\subsection*{Rigorous Error and Performance Analysis}

We analyse the truncation error associated with the standard fully-implicit finite-difference method and the particle method.  As such, let $\myhbar$ denote the exact solution of Equation~\eqref{eq:particle}, and let $\myhbar_{\Delta x}$ denote the numerical solution with step size $\Delta x$ (finite-difference method), or number of particles $N=2L/\Delta x$ (particle method). We assume that the error $\|\myhbar-\myhbar_{\Delta x}\|$ depends smoothly on $\Delta x$, then
\begin{equation}
    \|\myhbar-\myhbar_{\Delta x}\| = C\Delta x^p + O(\Delta x^{p+1}),
\end{equation}
for some constant $C$ and $p$. Since $\myhbar$ is unknown, we instead compute
\begin{equation}
\varepsilon(\Delta x) := \|\myhbar_{\Delta x}-\myhbar_{\Delta x/2}\|
\end{equation}
Using the triangle inequality, it can be shown that
%
\begin{equation}
    \varepsilon(\Delta x) \leq  C\Delta x^p(1-1/2^p) + O(\Delta x^{p+1}).
\label{eq:varepsilondef1}
\end{equation}
We take the natural logarithm on both sides of Equation~\eqref{eq:varepsilondef1}; this gives:
\begin{equation}
    \log\varepsilon \leq p\log(\Delta x) + \log(C) + \log(1-1/2^p).
\label{eq:varepsilondef2}
\end{equation}
Thus, the rate of convergence (or the order of accuracy) of the numerical method is $p$; $p$ can be computed from the numerical simulations as the slope of the log-log plot between the error $\varepsilon$ and the grid spacing $\Delta x$.

\begin{figure} [h]
    \begin{minipage}{0.45\textwidth}
        \centering
        \includegraphics[width=\linewidth]{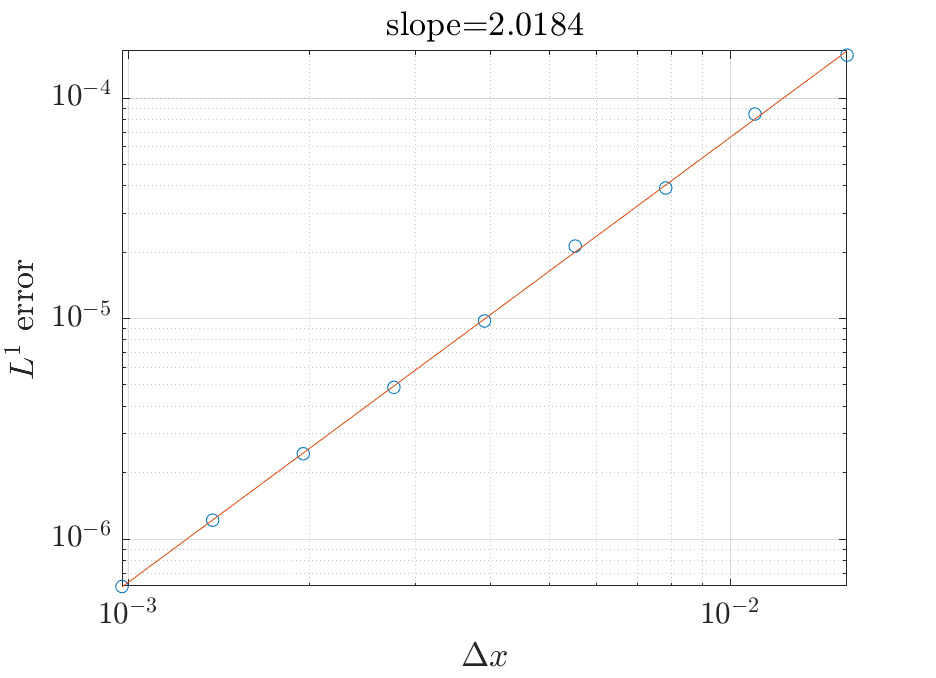}
    \end{minipage}
    \begin{minipage}{0.45\textwidth}
        \centering
        \includegraphics[width=\linewidth]{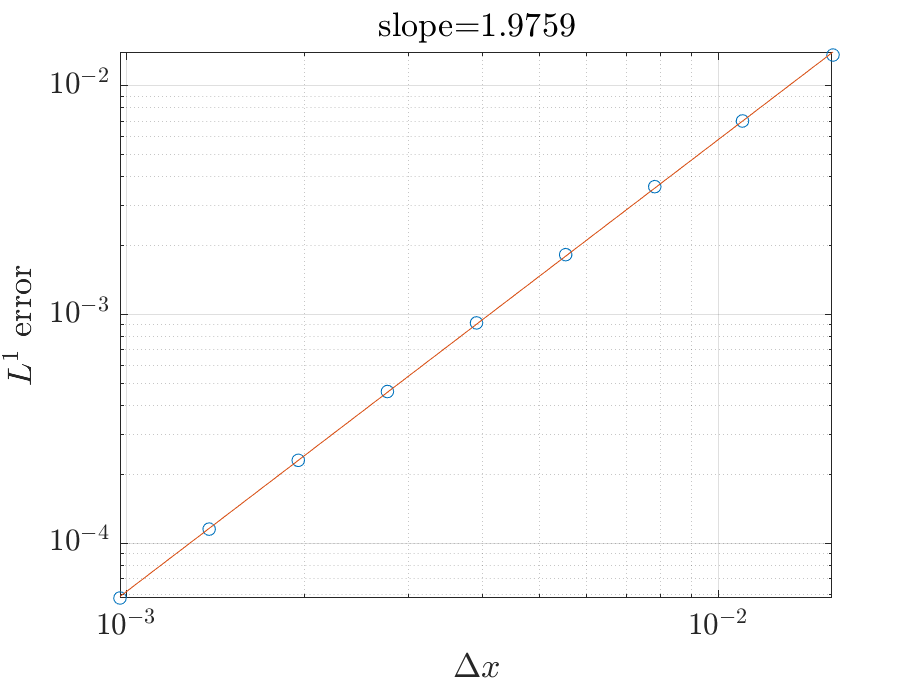}
    \end{minipage}
    \caption{Convergence plot of the finite-difference method (left) and the particle method (right).}
    \label{fig:convergence}
\end{figure}

Figure~\ref{fig:convergence} shows the rate of convergence of the finite-difference method and the particle method -- here we use the $L^1$ norm applied to Equation~\eqref{eq:varepsilondef1}--\eqref{eq:varepsilondef2} (our choice of norm is obtained because the particle solutions are expected to converge weakly in an $L^1$ function space, see Reference~\cite{chertock2012convergence}).   The 
finite-difference method is implemented with a step size of $\Delta t=0.01$, while the particle method uses the ODE45 solver in Matlab, hence an adaptive time step. Both methods use the same numerical parameter of, $\alpha=0.05$, with final time $T=1$, and periodic boundary condition on the spatial domain $x\in[-1,1]$. From this figure, both the particle method and the standard finite-difference method are estimated to be second-order accurate in the spatial domain.

\begin{figure} [h]
    \centering
    \includegraphics[width=0.6\linewidth]{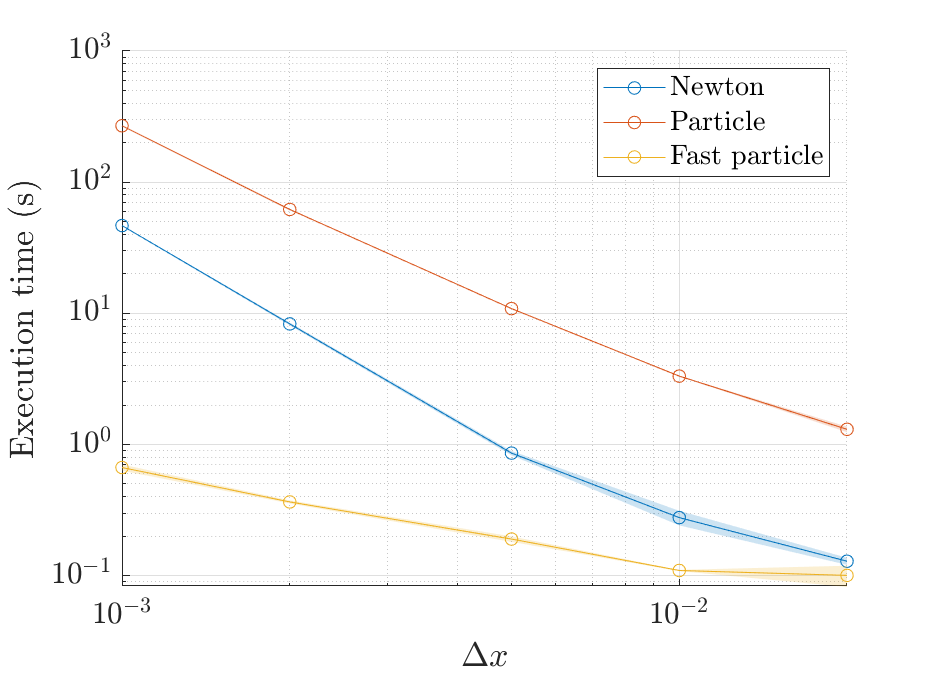}
    \caption{(a) Performance of the finite-difference method, the direct implementation of the particle method, and the fast implementation of the particle method.}
    \label{fig:performance}
\end{figure}

Furthermore,  in Figure~\ref{fig:performance} we evaluate the execution time of the different numerical methods to see if any one method outperforms the rest.   A comparison of the average execution time over 10 runs between the finite-difference method, the direct implementation of the particle method (computational complexity $O(N^2)$), and the fast implementation particle method (computational complexity $O(N)$). The numerical parameters used are the same as the one used in the convergence analysis and the calculations are performed on an Intel i7-9750H with 6 hyper-threaded cores. 
 The fast particle method is already comparable to the standard finite-difference method in terms of accuracy; from Figure~\ref{fig:performance} the fast particle method is seen to outperform the other methods, meaning that overall, the fast-particle method is the best method for simulating the droplet-spreading phenomenon.

\section{Partial Wetting}
\label{sec:partial}

In this section, we extend the Geometric Thin-Film Equation to the case of partial wetting, where a droplet on a substrate spreads initially before assuming an equilibrium shape.  This requires the addition of an extra, stabilizing term, to Equation~\eqref{eq:particle}.  We derive this additional term.  Then, we construct an analytical solution for the equilibrium droplet shape.  Finally, we use both the finite-difference method and the particle method to simulate transient droplet spreading, up to the point where the droplet assumes its equilibrium shape.

\subsection*{Theoretical Formulation}

The starting-point for the theoretical formulation is to consider an unregularized description of the droplet, with $h(x)$ as the droplet profile.  Then, the unregularized energy associated with a droplet of radius $r$ is:
\begin{equation}
\Energy=\gamma_{la}\int_{-r}^r \sqrt{1+ h_x^2}\mathd x+2r\,\gamma_{ls}+\gamma_{as}\left(S-2r\right),
\label{eq:physics}
\end{equation}
Here, $\gamma_{la}$ is the surface tension between the air and the liquid droplet (previously referred to as $\gamma$), $\gamma_{ls}$ is the surface tension between the liquid and the substrate, and $\gamma_{as}$ is the surface tension between the air and the substrate; $S$ is an arbitrary lengthscale denoting the extent of the system in the lateral direction.  In the longwave limit, $\sqrt{1+h_x^2}$ is expanded as $1+(1/2)h_x^2$, and Equation~\eqref{eq:physics} becomes:
\begin{equation}
\Energy=\tfrac{1}{2}\gamma_{la}\int_{-r}^r h_x^2\,\mathd x+2r\left(\gamma_{la}+\gamma_{ls}-\gamma_{as}\right)+\mathrm{Const.}
\label{eq:physics1}
\end{equation}
The three surface-tension coefficients are related via the Laplace-Young condition,
\begin{equation}
\gamma_{ls}+\gamma_{la}\cos\theta_{\mathrm{eq}}-\gamma_{as}=0,
\end{equation}
where $\thetaeq$ is the equilibrium contact angle.  Thus, Equation~\eqref{eq:physics1} can be re-written as
\begin{equation}
\Energy=\tfrac{1}{2}\gamma_{la}\int_{-r}^r h_x^2\,\mathd x+2r\gamma_{la}\left(1-\cos\theta_{\mathrm{eq}}\right)+\mathrm{Const.}
\label{eq:physics2}
\end{equation}
Now, inspired by the replacements $h\rightarrow \myhbar$ in Section~\ref{sec:theory}, we propose herein a regularized energy,
\begin{equation}
\Energy=\tfrac{1}{2}\gamma_{la}\int_{-\infty}^\infty h_x\myhbar_x \mathd x+\gamma_{la}w\left(1-\cos\theta_{\mathrm{eq}}\right)+\text{Const.},
\label{eq:physics3}
\end{equation}
where $w$ is an estimate of the size of the droplet footprint, based on the interfacial profile $h$, and on the smoothened interfacial profile, $\myhbar=\Phi*h$. 

We estimate the size of the droplet footprint as
\begin{equation}
w=c\frac{\|\myhbar\|_1^2}{\langle h,\myhbar\rangle }=c\frac{A_0^2}{\langle h,\myhbar\rangle}
\label{eq:wapprox}
\end{equation}
where $A_0$ is the constant droplet volume, $A_0=\int_{-\infty}^\infty h(x,t)\mathd x=\int_{-\infty}^\infty \myhbar(x,t)\mathd x$, and $c$ is an $O(1)$ parameter to be determined.
The estimate in Equation~\eqref{eq:wapprox} is dimensionally correct, but also yields good agreements with some model droplet profiles: for instance, if $h$ were a spherical cap, $h(x)=\max\{0,(3A_0/4r)[1-(x/r)^2]\}$, then we would have (by direct calculation) $\langle h,\myhbar\rangle=3A_0^2/5r + O(\alpha^2)$, and
\[
\frac{\|\myhbar\|_1^2}{\langle h,\myhbar\rangle}=\tfrac{5}{6}(2r) + O(\alpha^2)
\]
i.e. a width proportional to the droplet footprint $2r$, with a constant of proportionality $6/5 + O(\alpha^2)$ close to one.  
%
%
%
Thus, the regularized energy becomes:
\begin{equation}
\Energy=\tfrac{1}{2}\gamma\int_{-\infty}^\infty h_x \myhbar_x \,\mathd x+\gamma \chi \frac{A_0^2}{\langle h,\myhbar\rangle}.
\label{eq:E_partial}
\end{equation}
where $\chi=c(1-\cos\theta_{\mathrm{eq}})$ is an $O(1)$ constant, which will be selected \textit{a priori} in what follows; we also use $\gamma$ instead of $\gamma_{la}$, for consistency with the previous sections.  Finally, the constant term in the energy has been dropped in Equation~\eqref{eq:E_partial}, because only energy differences are important for the purpose of deriving evolution equations.

To derive the evolution equation associated with Equation~\eqref{eq:E_partial}, we use the framework of Darcy's Law introduced previously in Section~\ref{sec:theory}.  Thus, the generalized force associated with Equation~\eqref{eq:E_partial} is:
\[
f=-\frac{\partial}{\partial x}\frac{\delta \Energy}{\delta h},
\]
and the Darcy velocity is therefore $U=Mf$, where $M$ is the mobility; we again take $M=(1/3\mu)\myhbar^2$.  Thus, 
\[
f=-\frac{\partial}{\partial x}\left(-\gamma\partial_{xx}\myhbar-2\gamma \chi\frac{A_0^2}{\langle h,\myhbar\rangle^2}\myhbar\right).
\]
The evolution equation for $h$ which conserves $\int_{-\infty}^\infty h\mathd x$ is thus:
\[
\frac{\partial h}{\partial t}+\frac{\partial}{\partial x}\left(hU\right)=0,
\]
hence
\begin{equation}
\frac{\partial h}{\partial t}=\frac{\partial}{\partial x}\left[hM\frac{\partial}{\partial x}\left(-\gamma\myhbar_{xx}-2\gamma \chi \frac{A_0^2}{\langle h,\myhbar\rangle^2}\myhbar\right)\right].
\label{eq:particle_partial_dim}
\end{equation}

\subsection*{Non-dimensionalization}

We non-dimensionalize Equation~\eqref{eq:particle_partial_dim} using the lengthscale $h_0=\sqrt{A_0 \tan\theta_{\mathrm{eq}}}$ in the vertical direction and $L=\sqrt{A_0/\tan\theta_{\mathrm{eq}}}$ in the lateral direction.  Thus, $h$ is made dimensionless on $h_0$, $x$ is made dimensionless on $L$, and time is made dimensionless on the capillary timescale $(3\mu L/\gamma)(\tan^3\theta_{\mathrm{eq}})$.
In dimensionless variables, Equation~\eqref{eq:particle_partial_dim} now reads:
\begin{equation}
\frac{\partial h}{\partial t}=-\frac{\partial}{\partial x}\left[h\myhbar^2\frac{\partial}{\partial x}\left(\myhbar_{xx}+2\chi \frac{\myhbar}{\langle h,\myhbar\rangle^2}\right)\right].
\label{eq:particle_partial}
\end{equation}
Also, in dimensionless variables, $\int_{-\infty}^\infty h(x,t)\mathd x=1$.
Also in this context, the ratio $\epsilon=h_0/L$ is precisely $\tan\theta_{\mathrm{eq}}$; strictly speaking therefore, $\theta_{\mathrm{eq}}$ should be small, for inertial effects to be negligible, and hence, for the lubrication theory underlying Equation~\eqref{eq:particle_partial} to be valid.

\subsection*{Equilibrium Solution}

\begin{figure}
	\centering
		\includegraphics[width=0.6\textwidth]{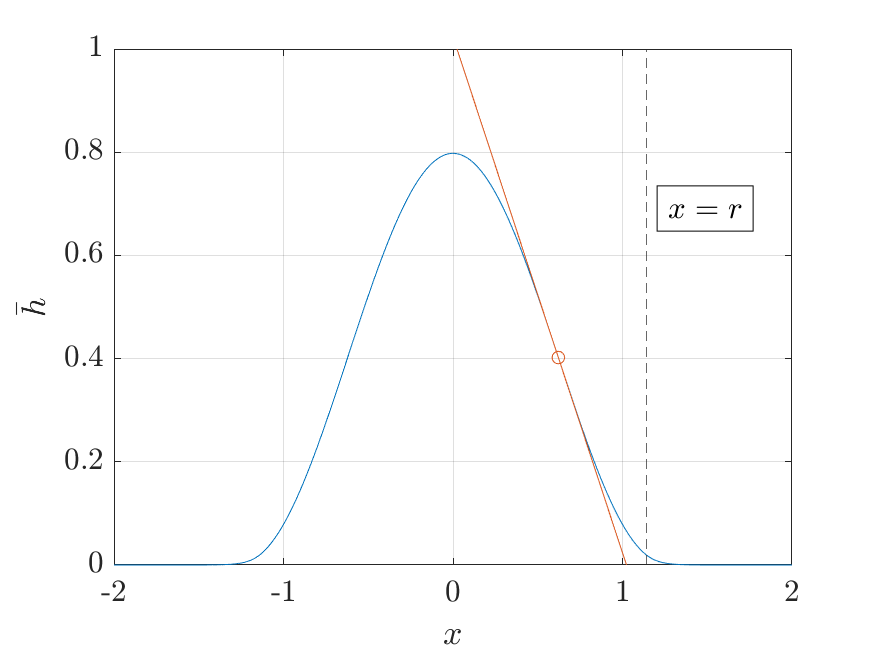}
		\caption{Equilibrium droplet profile $\myhbar(x)$ in dimensionless variables ($\alpha=0.05$). The red line is the tangent of the droplet profile at $x=\mathrm{argmax}_x[-\partial_{x}\myhbar(x)]$.}
	\label{fig:eqm}
\end{figure}

Equation~\eqref{eq:particle_partial} has an equilibrium solution with $\partial h/\partial t=0$.  In this limiting case, Equation~\eqref{eq:particle_partial} reduces to
\begin{equation}
h\myhbar^2\partial_x\left(\partial_{xx}\myhbar+\xi^2\myhbar\right)=0,
\label{eq:eqm1}
\end{equation}
where $\xi^2$ is a positive constant,
\begin{equation}
\xi^2=\frac{2\chi}{\langle h,\myhbar\rangle^2}.
\label{eq:eqm2}
\end{equation}
Equation~\eqref{eq:eqm1} has a simple analytical solution, parametrized by  $\xi$, and by a radius $r$:
\begin{equation}
\myhbar(x)=\begin{cases} B_1\cos(\xi x)+B_2. & |x|<r,\\
                  C_1 \mathe^{-|x|/\alpha}+C_2|x|\mathe^{-|x|/\alpha}, & |x|>r.\end{cases}
\label{eq:hbar_solution}
\end{equation} 
%
Correspondingly,
\begin{equation}
h(x)=\begin{cases} B_1(1+\alpha^2\xi^2)^2\cos(\xi x)+B_2, & |x|<r,\\
                          0, & |x|>r.\end{cases}
\label{eq:h_solution}
\end{equation} 
Here, $B_1,B_2,C_1$, and $C_2$ are constants of integration.  These constants are fixed by imposing continuity of $\myhbar,\myhbar_x,\myhbar_{xx}$ at $x=r$, and also by imposing $\int_{-\infty}^\infty \myhbar\, \mathd x=1$.  These give four conditions in four unknowns.  Hence, $\{B_1,B_2,C_1,C_2\}$ are fixed in terms of $r$.  The value of $r$ is in turn fixed by imposing continuity of $\myhbar_{xxx}$ at $x=r$, this gives a simple rootfinding condition:
\begin{equation}
\tan(\xi r)=-\frac{2\alpha \xi}{1-\alpha^2\xi^2}.
\label{eq:r_val}
\end{equation}
The details of this calculation are provided in Appendix~\ref{sec:app:eqm}.  Equation~\eqref{eq:r_val} gives $r$ as a function of $\xi$.  However, $\xi$ is not arbitrary, but is instead fixed by its own rootfinding condition (Equation~\eqref{eq:eqm2}).  Although this procedure is  somewhat involved, the point remains: the Geometric Thin-Film Equation with partial wetting admits an analytical equilibrium solution in terms of elementary functions (via Equations~\eqref{eq:hbar_solution}--\eqref{eq:h_solution}).  Furthermore, the elementary solution~\eqref{eq:hbar_solution} for $\myhbar(x)$ coincides with the expression for a spherical-cap droplet in the core region $x\rightarrow 0$,
\[
\myhbar(x)\approx \left(B_1+B_2\right)-\tfrac{1}{2}B_1\xi^2x^2,\qquad x\rightarrow 0.
\]
The equilibrium contact angle is now computed as:
\[
\tan\theta_{\mathrm{eq}}=-\epsilon \myhbar_x(x=x_{\mathrm{ref}}),
\]
where  $\myhbar_x$ is expressed in dimensionless variables and $x_{\mathrm{ref}}>0$ is a reference point.  Since $\tan\theta_{\mathrm{eq}}=\epsilon$ in the chosen dimensionless variables, this requires:
\[
1=-\myhbar_x(x=x_{\mathrm{ref}}),
\]
We choose the reference point $x_{\mathrm{ref}}$ to be the positive value of $x$ which maximizes $|\myhbar_x|$, thus $x_{\mathrm{ref}}=\pi/\xi$.  Thus, we require $B_1\xi=1$.  But $B_1$ and $\xi$ depend parametrically on $\chi$, hence we require $B_1(\chi)\xi(\chi)=1$.  This therefore fixes the model parameter $\chi$ as a global constant, $\chi\approx 1.1602$ (for filter width $\alpha=0.05$).  The resulting equilibrium droplet profile is shown in Figure~\ref{fig:eqm}.  Values of $\chi$ for different values of $\alpha$ are given in Table~\ref{tab:opt}.

\begin{table}[htb]
	\centering
		\begin{tabular}{|c|c|}
		\hline
		$\alpha$& $\chi$ \\
		\hline
		\hline
		0.01   & 1.1264\\
		0.02   & 1.1306\\
		0.05   & 1.1602\\
		\hline	
		\end{tabular}
		\caption{Optimum value of $\chi$ for different values of filter width $\alpha$}
		\label{tab:opt}
\end{table}


\subsection*{Results}

\begin{figure}[htb]
	\begin{minipage}{0.45\textwidth}
        \centering
        \includegraphics[width=\linewidth]{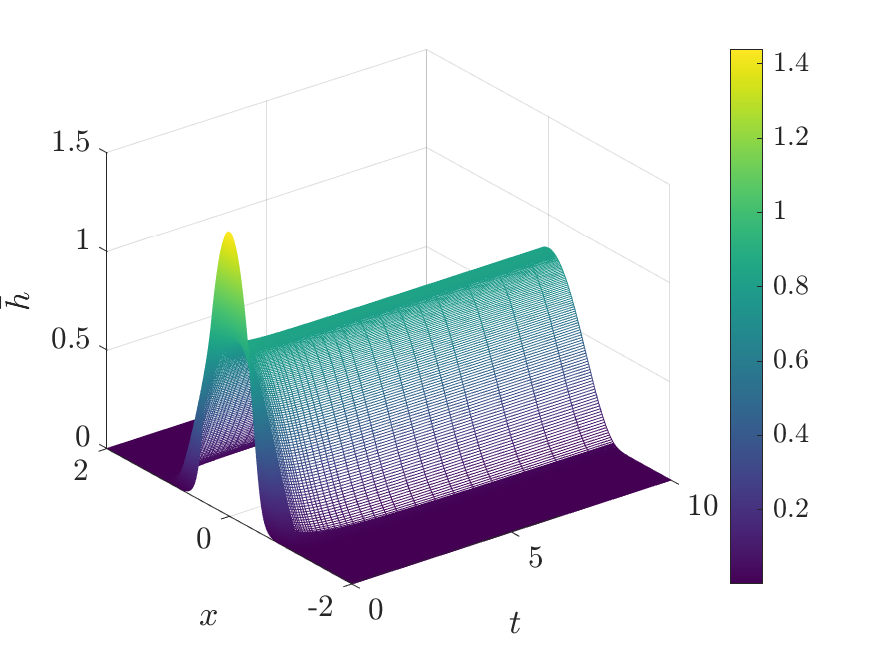}
    \end{minipage}
    \begin{minipage}{0.45\textwidth}
        \centering
        \includegraphics[width=\linewidth]{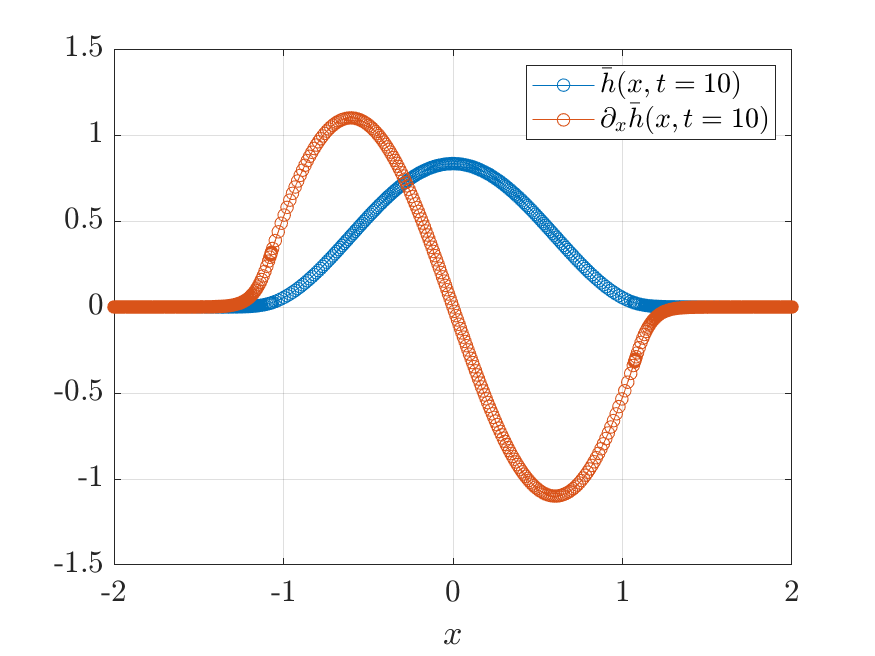}
    \end{minipage}
    \caption{(a) Space-time plot of $\myhbar(x,t)$ showing the spreading of the droplet for the partial-wetting case. (b) Droplet shape $\myhbar$ and the slope $\partial_x\myhbar$ at $t=10$. 
		}
    \label{fig:eq_solution}
\end{figure}

\begin{figure}
	\centering
		\includegraphics[width=0.6\textwidth]{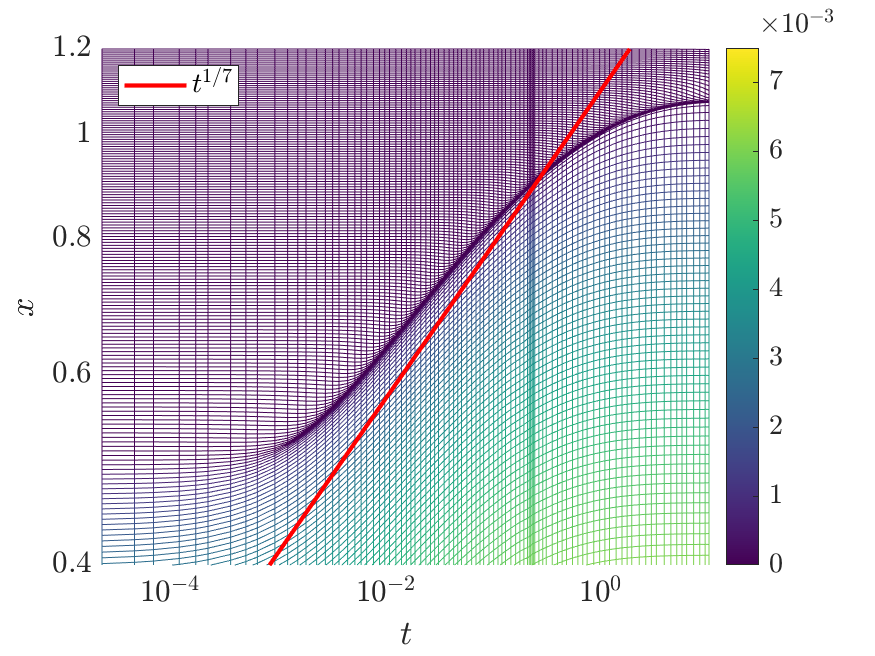}
		\caption{Spacetime plot showing the evolution of the particle trajectories (partial-wetting).  Logarithmic scale on both axes.}
	\label{fig:eq_trajectory}
\end{figure}

For the simulations of partial wetting, we use the initial condition
\[
h(x,t=0)=\begin{cases}\frac{3}{4r_0}\left[1-(x/r_0)^2\right],&|x|<r_0,\\
                      0,&|x|>r_0.\end{cases}
\]
with $r_0=0.5$, and $\int_{-\infty}^\infty h(x,t=0)\,\mathd x=1$.  We also take $\alpha=0.05$.  We use both the particle method and the finite-difference method: the results are the same in each case.  In Figure~\ref{fig:eq_solution}(a), we show a space-time plot of the solution for the partial wetting case up to $t=10$; panel (b) shows a snapshot of the droplet profile at $t=10$. Figure~\ref{fig:eq_trajectory} is based exclusively on the particle method: here we show a log-log plot of the particle trajectories.  At intermediate times, the particle trajectories are parallel to the path $x=t^{1/7}$ before attaining a steady state at late times.  Thus, in the case of partial wetting, the system obeys Tanner's law at intermediate times, until at late times, the partial wetting stabilizes the droplet and it assumes its equilibrium shape.


\begin{figure}
	\centering
		\includegraphics[width=0.6\textwidth]{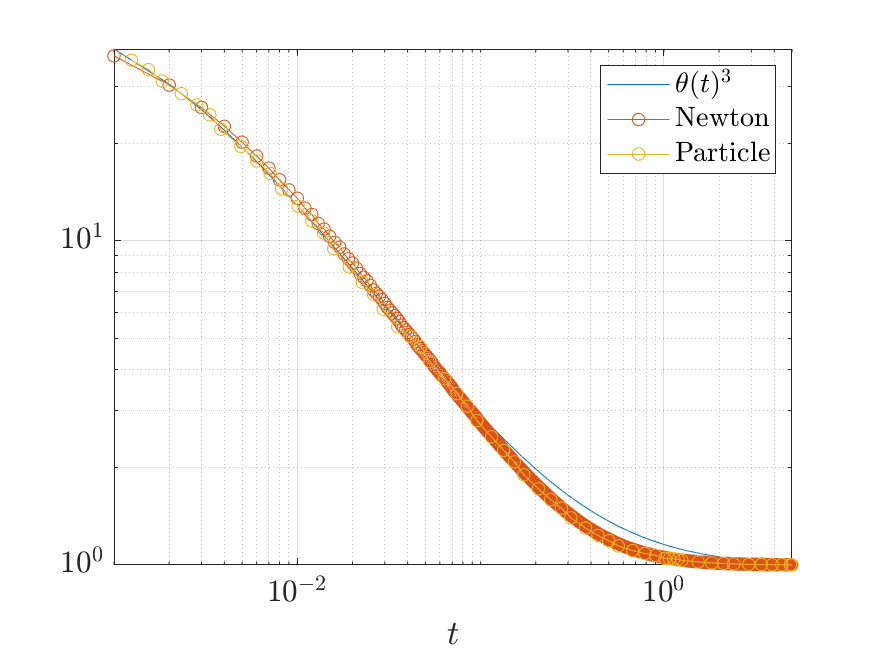}
		\caption{Plot of $[\theta(t)]^3$ (solid line) and $1+c\dot{x}_{cl}\log(x_{cl}/d)$ (solid line with markers) as a function of time showing the agreement between GDIM and the Cox--Voinov theory for droplet spreading in the case of partial wetting.  Here, the dynamic contact angle $\theta(t)$ and the contact-line position $x_{cl}(t)$ are defined operationally as in the text.  The values of $c$ and $d$ are chosen to optimize the fit between the two curves.}
	\label{fig:cox_voinov2}
\end{figure}

The foregoing statement that the droplet spreading obeys Tanner's Law at intermediate times until the onset of equilibrium also applies to the Cox--Voinov Law~\cite{bonn2009wetting} of droplet spreading, which in equation form is $[\theta(t)]^3=[\theta_{\mathrm{eq}}]^3+c\dot{x}_{\mathrm{cl}}\log(x_{\mathrm{cl}}/d)$.  Here, $\theta(t)$ is the dynamic contact angle, which is obtained from the slope of the interface profile $h(x,t)$ at some appropriate location $x$, and $c$ and $d$ are constants.  In order to validate the applicability of the Cox--Voinov law to the droplet spreading in GDIM, we operationally define the contact angle as $\theta(t)=\max_x [-\partial_x\myhbar(x,t)])$.  The tangent line to $\myhbar(x,t)$ at $\mathrm{argmax}_x[-\partial_x\myhbar(x,t)]$ is constructed, and the contact line is then taken to be the intersection of this tangent line with the $x$-axis.  A plot of $[\theta(t)]^3$ constructed in this way is shown in Figure~\ref{fig:cox_voinov2}.  Shown also is a plot 
$1+c\dot{x}_{\mathrm{cl}}\log(x_{\mathrm{cl}}/d)$ -- here, $\theta_{\mathrm{eq}}=1$, and $c$ and $d$ are best-fit constants.  Overall, there is good agreement between the two curves at intermediate times -- consistent with the behaviour of the trajectories Figure~\ref{fig:eq_solution}.  There is some disagreement at late times, however, this may be expected, in view of the somewhat imprecise operational definition of $\theta(t)$ and $x_{\mathrm{cl}}$.

\subsection*{Rigorous Error and Performance Analysis}

\begin{figure} [h]
    \begin{minipage}{0.45\textwidth}
        \centering
        \includegraphics[width=\linewidth]{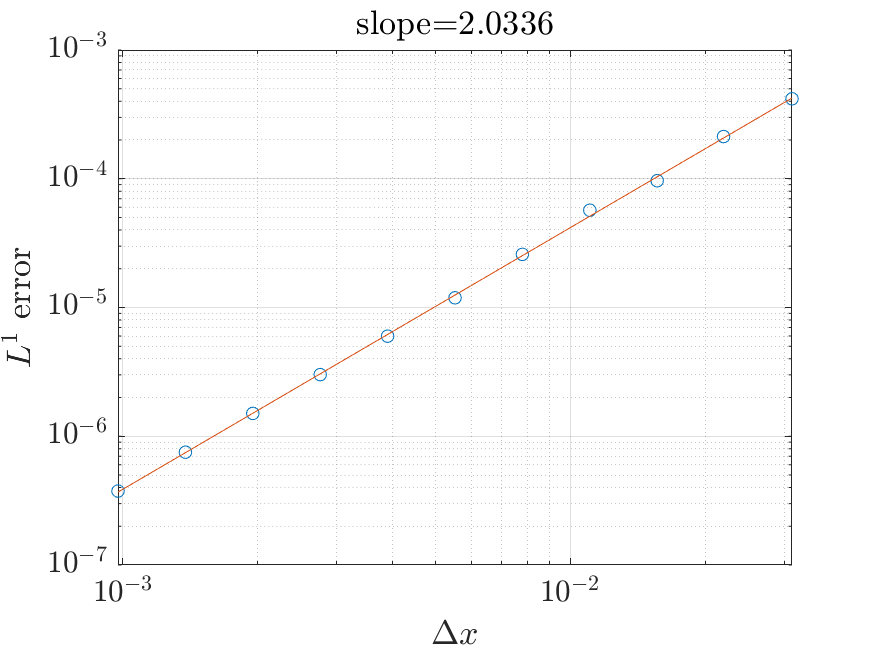}
    \end{minipage}
    \begin{minipage}{0.45\textwidth}
        \centering
        \includegraphics[width=\linewidth]{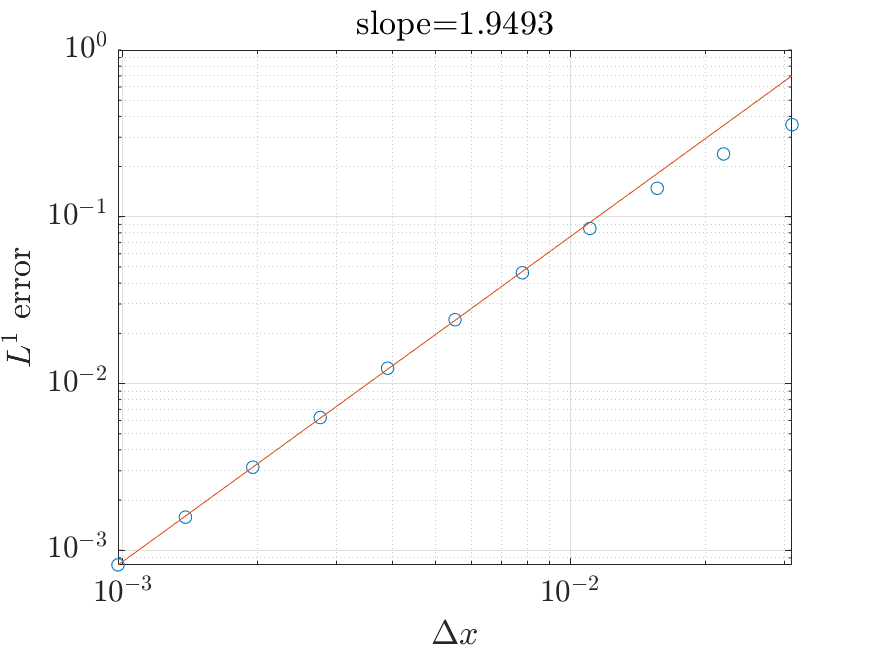}
    \end{minipage}
    \caption{Convergence plot for the partial wetting case of the finite-difference method (left) and the particle method (right).}
    \label{fig:eq_convergence}
\end{figure}

We carry out a rigorous error analysis of both the fully-implicit finite-difference method, and the particle method, for the case of partial wetting.  Because there is an analytical, equilibrium solution valid at late times, we analyze the results of the numerical simulations at such late time (specifically, $t=100$), when the numerical solutions attain equilibrium.  In this case, the
 the equation for the rate of convergence of the numerical method is simply
\begin{equation}
\log\|\myhbar-\myhbar_{\Delta x}\|_1 = p\log(\Delta x),
\end{equation}
where $\myhbar$ denotes the analytical equilibrium solution, and $\myhbar_{\Delta x}$ denotes the numerical  equilibrium solution (or what amounts to the same, the numerical solution at $t=100$).
Figure~\ref{fig:eq_convergence} shows the rate of convergence for the finite-difference method and the particle method. Both methods have $\alpha=0.05$ and were performed on the spatial domain $x\in[-2,2]$ for various spatial resolutions $\Delta x$.  We have used the MATLAB solver ODE15s for the particle method. Again, we observed both the finite-difference method and the particle method to be second-order accurate in the spatial domain. 
%

\begin{figure}[htb]
    \centering
    \includegraphics[width=0.6\linewidth]{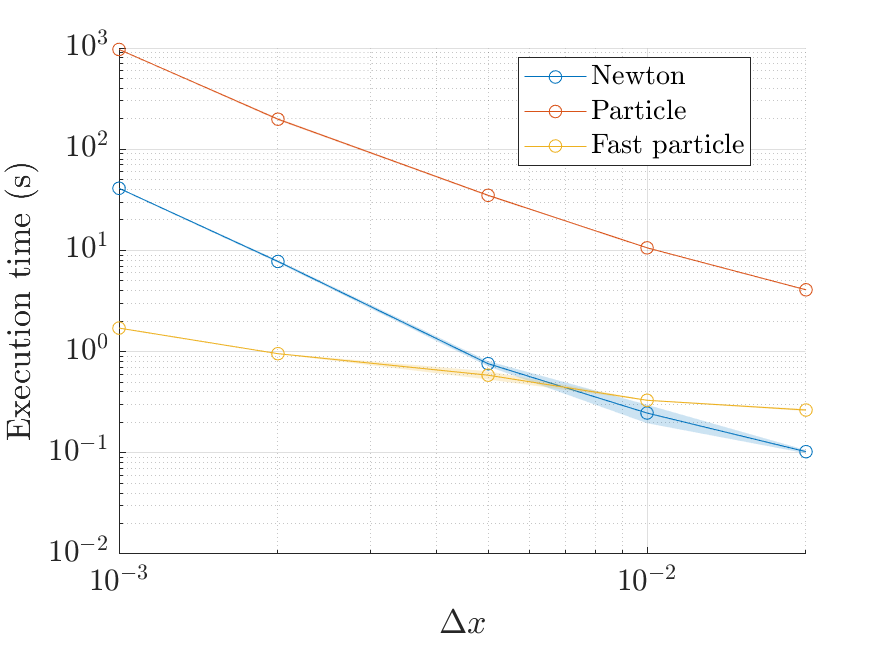}
    \caption{Performance of the finite-difference method, the direct implementation of the particle method, and the fast implementation of the particle method (partial wetting)}
    \label{fig:perf_eq}
\end{figure}

Finally, we have looked at the performance of the different numerical methods (fully-implicit finite-difference method, particle method, and `fast' particle method): the results are similar to what was observed in the case of complete wetting (Figure~\ref{fig:perf_eq}).

\section{Conclusions}
\label{sec:conclusions}

Summarizing, we have introduced a new mathematical model to describe contact-line motion which has the same regime of validity as conventional lubrication theory.  The model involves using a `smooth'
interface profile $\myhbar$ and a sharp interface profile $h$.  The smooth interface profile $\myhbar$ is connected to the sharp interface profile $h$ via convolution, while $h$ is defined through an evolution equation which couples both interface profiles, and which drives the droplet spreading.  It is possible to assign a physical interpretation to the model: the equation for $h$ is a model with missing small-scale physics (below a lengthscale $\alpha$), while $\myhbar$ is a complete model of the physics at lengthscales greater than $\alpha$.  By minimizing the difference between the two descriptions, a convolution relation between $\myhbar$ and $h$ emerges.  Furthermore, by formulating the evolution equation so that it involves both descriptions of the interface profile, the contact-line singularity is resolved.  There is no need to model the missing small-scale physics explicitly: instead, it is parametrized through the lengthscale $\alpha$.

A further advantage of the new formulation is that it naturally suggests a mesh-free numerical method for the purpose of simulating the model numerically.  Furthermore, the mathematical model involves equations which are non-stiff, and are straightforward to solve numerically (we include a repository of the code as part of this work; see Reference~\cite{Github}).    We have conducted numerical simulations for both complete wetting and partial wetting using this new mesh-free method (as well as a conventional finite-difference method), and have found that the model describes well the physics of droplet spreading -- including Tanner's Law for the evolution of the contact line.  Remarkably, in the case of partial wetting, the model also admits a simple analytical solution for the equilibrium profile.

Beyond droplet spreading, the model may find further applications in describing families of droplets (the analytical equilibrium solution already possesses such `multiple-droplet' solutions), multi-component systems, and problems in droplet evaporation.  The model's intrinsically non-stiff equations (as well as the absence of a precursor film extending to infinity) may in future simplify the description of such complex physical systems.

\subsection*{Acknowledgements}

This publication has emanated from research supported in part by a Grant from Science Foundation Ireland under Grant number 18/CRT/6049.
LON has also been supported by the ThermaSMART network. The ThermaSMART network has received funding from the European Union’s Horizon 2020 research and innovation programme under the Marie Sklodowska--Curie grant agreement No. 778104.  

\appendix

\section{Detailed Description of Numerical Methods}
\label{sec:app:numerics}

In this Appendix, we describe in detail the numerical methods used in the paper.  For definiteness, we focus on the case of complete wetting, although the description carries over to the case of partial wetting as well.  We start by describing the fully-implicit finite-difference method, and then describe the fast-particle method.

\subsection*{Fully-Implicit Finite-Difference Method}

Instead of solving Equation~\eqref{eq:particle}, we instead take the convolution of both sides of the same equation with the filter function $\Phi$ and solve the evolution for the smoothened free-surface height $\myhbar(x,t)$:
\begin{equation} 
	\frac{\partial \myhbar}{\partial t} = -\Phi*\left[\frac{\partial}{\partial x}\left(h\myhbar^2 \partial_{xxx}\myhbar\right)\right],\qquad \myhbar=\Phi*h.
	\label{eq:smooth_tfe}
\end{equation}
Here, we have written $K*K=\Phi$ ($K$ is the Helmholtz kernel) such that $\myhbar=\Phi*h$, reflecting the choice of the inverse-double-Helmholtz operator as the smoothing kernel in Equation~\eqref{eq:particle}.  We also work in dimensionless variables, such that the prefacor $\surften/(3\mu)$ which was present in Equation~\eqref{eq:particle} is rescaled to one.
Equation~\eqref{eq:smooth_tfe} is discretized in space using a standard finite-difference method.  The derivatives are approximated using standard finite-difference stencils (central differencing, first and second-order derivatives second order accurate in space).  The value of $\myhbar$ is therefore known only at the discrete grid points; we collect these values into a vector $\bar{\bm{h}}$.  Similarly, Equation~\eqref{eq:smooth_tfe} is discretized in time such that the values of $\myhbar$ are known only at discrete points in time; the fixed interval between each such discrete point in time is the timesetep $\Delta t$.  The values of the vector $\bar{\bm{h}}$ at a particular discrete point in time are therefore labelled as $\bar{\bm{h}}^n$.  Thus, in discrete form, Equation~\eqref{eq:smooth_tfe} becomes: 
\begin{equation}
	\frac{\bar{\bm{h}}^{n+1}-\bar{\bm{h}}^n}{\Delta t} = -\mathcal{K}\mathcal{D}_1\left[\bm{h}^{n+1}\odot(\bar{\bm{h}}^{n+1})^2\odot(\mathcal{D}_3\bar{\bm{h}}^{n+1})\right],\qquad \bm{h}^{n+1}=\mathcal{K}^{-1}\bar{\bm{h}}^{n+1}.
	\label{eq:discretize_tfe}
\end{equation}
Here, $\mathcal{D}_1$ and $\mathcal{D}_3$ are sparse matrices corresponding to the centred finite-difference approximation of the first and third spatial derivatives, respectively.  A similar notation would hold for $\mathcal{D}_2$, which corresponds to the centred finite-difference approximation of the second spatial derivative.  In this way, $\mathcal{K}$ is a convolution matrix, which is constructed as $\mathcal{K}=(1-\alpha^2\mathcal{D}_2)^{-2}$.   Furthermore, we use the $\odot$ notation in Equation~\eqref{eq:discretize_tfe} to denote the pointwise multiplication of vectors, e.g. if $\bm{a}$ and $\bm{b}$ are $N$-dimensional column vectors with entries $\bm{a}=(a_1,\cdots,a_N)^T$ and $\bm{b}=(b_1,\cdots,b_N)^T$, then $\bm{a}\odot\bm{b}$ is also an $N$-dimensional row vector with entries $\bm{a}\odot\bm{b}=(a_1b_1,\cdots,a_Nb_N)^T$.  This notation is useful for the exposition of the numerical method, as the numerical code for solving Equation~\eqref{eq:smooth_tfe} is vectorized, and makes use of precisely this pointwise vector multiplication.

Crucially, in the discretized equation~\eqref{eq:discretize_tfe} we have treated the high-order spatial derivative on the right-hand-side fully implicitly ($\bar{\bm{h}}^{n+1}$ instead of 
$\bar{\bm{h}}^{n}$).  A fully explicit treatment ($\bar{\bm{h}}^{n}$ instead of  $\bar{\bm{h}}^{n+1}$) would suffer from a severe hyperdiffusion-limited CFL constraint $\Delta t< C \Delta x^4$, where $\Delta x$ is the grid spacing, and $C$ is an $O(1)$ constant.  Although this makes the numerical code unconditionally stable, it means that at each time-step, a set of nonlinear equations for $\bar{\bm{h}}^{n+1}$ must be solved.  We therefore describe how these nonlinear equations are solved.

We introduce
\begin{equation}
	\bm{F}(\bar{\bm{v}}) = \bar{\bm{v}} + \Delta t \bar{\mathcal{D}}_1\left(\bm{v}\odot\bar{\bm{v}}^2\odot(\mathcal{D}_3\bar{\bm{v}})\right)-\bar{\bm{h}}^n,
\end{equation}
where $\bar{\mathcal{D}}_1=\mathcal{K}\mathcal{D}_1, \mathcal{D}_3=\mathcal{D}_3\mathcal{K}$, and $\bm{v} = \mathcal{K}^{-1}\bar{\bm{v}}$. Then, solving Equation~\eqref{eq:discretize_tfe} for $\bar{\bm{h}}^{n+1}$ given $\bar{\bm{h}}^n$ becomes a minimization problem on 
\begin{equation}
	f(\bar{\bm{h}}^{n+1}) = \frac{1}{2}\|\bm{F}(\bar{\bm{h}}^{n+1})\|.
\end{equation}
We employ a Newton-Linesearch method to solve for $\bar{\bm{h}}^{n+1}$. This requires us to compute the Jacobian of $\bm{F}(\bar{\bm{v}})$, which is given by
\begin{equation}
	J(\bar{\bm{v}}) = I + \Delta t \left((\bar{\bm{v}}^2\odot \mathcal{K}^{-1} + 2\bar{\bm{v}}\odot\bm{v}) \odot \mathcal{D}_3\bar{\bm{v}} + (\bm{v}\odot\bar{\bm{v}}^2)\odot\mathcal{D}_3 \right).
\end{equation}
We initialize the guess with $\bar{\bm{v}}^0=\bar{\bm{h}}^n$. Then, for each iteration $I$, the descent direction is given by
\begin{equation}
	\delta\bar{\bm{v}}^I = -\left(J(\bar{\bm{v}}^I)^{-1}\bm{F}(\bar{\bm{v}}^I)\right),
\end{equation}
and the improved guess is updated using
\begin{equation}
	\bar{\bm{v}}^{I+1} = \bar{\bm{v}}^I + \alpha^I\delta\bar{\bm{v}}^I,
\end{equation}
where $\alpha^I$ is the optimum step size in the direction $\delta\bar{\bm{v}}^I$
\begin{equation}
	\alpha^I = \argmin\limits_{\alpha\in[0,1]}\left\{f\left(\bar{\bm{v}}^I+\alpha\delta\bar{\bm{v}}^I\right)\right\}.
\end{equation}
The iterative process is continued until the residual $f(\bar{\bm{v}}^I)$ is sufficiently small, and we set the solution for the next time step by letting $\bar{\bm{h}}^{n+1} = \bar{\bm{v}}^I$.

\subsection*{Fast summation algorithm for particle method}

Here, we are interested in solving the Geometric Thin-Film equation~\eqref{eq:particle} using the particle method.  The  particle method requires us to compute the evolution equations for the particle trajectories, these are given in Equation~\eqref{eq:ode}.  To solve Equation~\eqref{eq:ode},  one is required to evaluate
\begin{align}
	\myhbar^N(x_i) &= \sum_{j=1}^N w_j \Phi(x_i-x_j), \label{eq:particle_sum1} \\
	\partial_{xxx}\myhbar^N(x_i) &= \sum_{j=1}^N w_j \Phi'''(x_j-x_i), \label{eq:particle_sum2}
\end{align}
for $i=1,\dots,N$, at each time step. A direct evaluation would require an operational cost of $O(N^2)$. Now, suppose that $x_i<x_j$ for all $i<j$, then the contribution from particles of either side of $x_i$ becomes a degenerate case of the fast multipole method. We start by defining new variables
\begin{equation}
S_1 = \sum_{j=1}^{i-1}w_j\Phi(x_i-x_j), \qquad S_2 = \sum_{j=i+1}^{N}w_j\Phi(x_i-x_j).
\end{equation}
Since $x_i-x_j>0$ for $j=1,\dots,i-1$, we can drop the absolute value in the function $\Phi$,
\begin{align}
    S_1 &= \sum_{j=1}^{i-1}w_j(\alpha+x_i-x_j)\mathe^{-(x_i-x_j)/\alpha}, \\
    &= (\alpha+x_i)\mathe^{-x_i/\alpha}\underbrace{\sum_{j=1}^{i-1}w_j\mathe^{x_j/\alpha}}_{a^{(0)}_i} - \mathe^{-x_i/\alpha}\underbrace{\sum_{j=1}^{i-1}w_jx_j\mathe^{x_j/\alpha}}_{a^{(1)}_i}.
\end{align}
Similarly, we obtain for $S_2$,
\begin{align}
    S_2 &= (\alpha-x_i)\mathe^{x_i/\alpha}\underbrace{\sum_{j=i+1}^Nw_j\mathe^{-x_j/\alpha}}_{b^{(0)}_i} + \mathe^{x_i/\alpha}\underbrace{\sum_{j=i+1}^Nw_jx_j\mathe^{-x_j/\alpha}}_{b^{(1)}_i}.
\end{align}
So Equation~\eqref{eq:particle_sum1} can be expressed as 
\begin{align}
    \myhbar^N(x_i) &= S_1 + w_i\Phi(0) + S_2, \\
    &= \frac{1}{4\alpha^2}\mathe^{-x_i/\alpha}((\alpha+x_i)a^{(0)}_i-a^{(1)}_i) + w_i\Phi(0) \nonumber \\ 
    &\qquad + \frac{1}{4\alpha^2}\mathe^{x_i/\alpha}((\alpha-x_i)b^{(0)}_i+b^{(1)}_i).
\end{align}
Note that $a^{(0)}_i, a^{(1)}_i, b^{(0)}_i, b^{(1)}_i$ can be computed before hand using a recursion formula
\begin{align}
&
\begin{rcases}
    a^{(0)}_1=0, & \qquad a^{(0)}_i = a^{(0)}_{i-1} + w_{i-1}\mathe^{x_{i-1}/\alpha} \\
    a^{(1)}_1=0, & \qquad a^{(1)}_i = a^{(1)}_{i-1} + w_{i-1}x_{i-1}\mathe^{x_{i-1}/\alpha}
\end{rcases}
\ i=2,\dots,N, \\ 
&
\begin{rcases}
    b^{(0)}_N=0, & \qquad b^{(0)}_i = b^{(0)}_{i+1} + w_{i+1}\mathe^{-x_{i+1}/\alpha} \\
    b^{(1)}_N=0, & \qquad b^{(1)}_i = b^{(1)}_{i+1} + w_{i+1}x_{i+1}\mathe^{-x_{i+1}/\alpha}
\end{rcases}
\ i=N-1,\dots,1.
\end{align}
We follow the same procedure for Equation~\eqref{eq:particle_sum2} to obtain, 
\begin{align}
    \partial_{xxx}\myhbar^N(x_i) &= \frac{1}{4\alpha^3}\mathe^{-x_i/\alpha}((2\alpha-x_i)a_i^{(0)}+a_i^{(1)}) \nonumber\\
    &\qquad + \frac{1}{4\alpha^3}\mathe^{x_i/\alpha}(-(2\alpha+x_i)b_i^{(0)}+b_i^{(1)}).
\end{align}
In the case of the partial wetting, the term $\partial_x\myhbar$ and $\langle h,\myhbar\rangle$ in Equation~\eqref{eq:particle_partial_dim} would be decomposed using the same procedure. This decomposition reduces the complexity of evaluating the Equation~\eqref{eq:ode} from $O(N^2)$ to $O(N)$.

\section{Derivation of the Equilibrium Solution for Partial Wetting}
\label{sec:app:eqm}

In this Appendix, we look at the equilibrium solution for the droplet in the case of partial wetting.  The equilibrium solution was outlined in Section~\ref{sec:partial}, and involved expressions for $h(x)$ and $\myhbar(x)$ in terms of elementary functions.  We begin by recalling these expressions:
\begin{subequations}
\begin{equation}
    h(x) =
    \begin{dcases}
    B_1(1+\alpha^2\xi^2)^2\cos(\xi x) + B_2 & \text{for $|x|\leq r$}, \\
    0 & \text{for $|x|\geq r$},
    \end{dcases} 
\end{equation}
\begin{equation}
    \myhbar(x) =
    \begin{dcases}
    B_1\cos(\xi x) + B_2 & \text{for $|x|\leq r$}, \\
    C_1\exp\left(-\frac{|x|}{\alpha}\right) + C_2|x|\exp\left(-\frac{|x|}{\alpha}\right) & \text{for $|x|\geq r$}.
    \end{dcases}
\end{equation}
\end{subequations}
Here, $B_1$, $B_2$, $C_1$, and $C_2$ (together with $r$) are constants of integration, in this Appendix we derive closed-for expressions for these constants.

We start by noticing that the set of boundary conditions 
\begin{enumerate}
    \item $A_0=\int h\,\mathd x$,
    \item $A_0=\int \myhbar\,\mathd x$,
    \item $\myhbar'$ continuous at $r$, and
    \item $\myhbar'''$ continuous at $r$, 
\end{enumerate}
are linearly \textit{dependent}. This can be seen by first writing out the boundary conditions 2, 3, and 4 explicitly:
\begin{gather}
    \frac{1}{2}A_0 = B_1\frac{1}{\xi}\sin(\xi r) + B_2r + \alpha(C_1+rC_2+\alpha C_2)\exp\left(-\frac{r}{\alpha}\right), \\
    -B_1\xi\sin(\xi r) = -\frac{1}{\alpha}(C_1-\alpha C_2+rC_2)\exp\left(-\frac{r}{\alpha}\right), \\
    B_1\xi^3\sin(\xi r) = -\frac{1}{\alpha^3}(C_1-3\alpha C_2+rC_2)\exp\left(-\frac{r}{\alpha}\right).
\end{gather}
Some rearranging gives
\begin{gather}
    \frac{1}{\alpha\xi}B_1\sin(\xi r) + B_2\frac{r}{\alpha} = \frac{1}{2\alpha}A_0 -(C_1+rC_2+\alpha C_2)\exp\left(-\frac{r}{\alpha}\right), \label{eq:linear_ind1}\\
    -\frac{1}{\xi\alpha}\xi^2\alpha^2B_1\sin(\xi r) = -(C_1-\alpha C_2+rC_2)\exp\left(-\frac{r}{\alpha}\right), \label{eq:linear_ind2}\\
    \frac{1}{\xi\alpha}\xi^4\alpha^4B_1\sin(\xi r) = -(C_1-3\alpha C_2+rC_2)\exp\left(-\frac{r}{\alpha}\right).
    \label{eq:linear_ind3}
\end{gather}
Now take $\eqref{eq:linear_ind1}-2\eqref{eq:linear_ind2}+\eqref{eq:linear_ind3}$,
\begin{align}
    \frac{1}{\alpha\xi}B_1(1+\alpha^2\xi^2)^2\sin(\xi r) + B_2\frac{r}{\alpha} &= \frac{1}{2\alpha}A_0, \\
    B_1(1+\alpha^2\xi^2)^2\frac{1}{\xi}\sin(\xi r) + B_2r &= \frac{1}{2}A_0,
\end{align}
which is the equation of Boundary Condition 1.
Therefore choosing any three out of the four boundary conditions yield the same system of equations.

Next, consider the following set of boundary conditions for expression (1)
\begin{gather}
    A_0 = \int h\,\mathd x = \int \myhbar\,\mathd x, \\
    \text{$\myhbar$, $\myhbar'$, and $\myhbar''$ continuous at $r$.} 
\end{gather}
This gives a system of five equations
\begin{gather}
    \frac{1}{2}A_0 = B_1(1+\alpha^2\xi^2)^2\frac{1}{\xi}\sin(\xi r) + B_2r, \label{eq:bc1}\\
    \frac{1}{2}A_0 = B_1\frac{1}{\xi}\sin(\xi r) + B_2r + \alpha(C_1+rC_2+\alpha C_2)\exp\left(-\frac{r}{\alpha}\right), \label{eq:bc2}\\
    B_1\cos(\xi r) + B_2 = (C_1+rC_2)\exp\left(-\frac{r}{\alpha}\right), \label{eq:bc3}\\
    -B_1\xi\sin(\xi r) = -\frac{1}{\alpha}(C_1-\alpha C_2+rC_2)\exp\left(-\frac{r}{\alpha}\right), \label{eq:bc4}\\ 
    -B_1\xi^2\cos(\xi r) = \frac{1}{\alpha^2}(C_1-2\alpha C_2+rC_2)\exp\left(-\frac{r}{\alpha}\right). \label{eq:bc5}
\end{gather}
Subtract Equation~\eqref{eq:bc1} from Equation~\eqref{eq:bc2} gives
\begin{align}
\frac{B_1}{\xi}\sin(\xi r)\left[(1+\alpha^2\xi^2)^2-1\right] &= \alpha(C_1+rC_2+\alpha C_2)\exp\left(-\frac{r}{\alpha}\right), 
\end{align}
then divide by Equation~\eqref{eq:bc4}
\begin{align}
-\frac{1}{\xi^2}\left((1+\alpha^2\xi^2)^2-1\right) &= -\alpha^2\frac{C_1+rC_2+\alpha C_2}{C_1-\alpha C_2+rC_2}, \\
2 + \alpha^2\xi^2 &= \frac{C_1+rC_2+\alpha C_2}{C_1+rC_2-\alpha C_2}.
\label{eq:linear_eq1}
\end{align}
To solve for $C_1$ and $C_2$, we need one more equation, for this, take $\eqref{eq:bc4} \div \eqref{eq:bc5}$
\begin{align}
\frac{1}{\xi}\tan(\xi r) &= -\frac{1}{\alpha}\frac{C_1+rC_2-\alpha C_2}{C_1+rC_2-2\alpha C_2}.
\label{eq:linear_eq2}
\end{align}
Then we can write Equation~\eqref{eq:linear_eq1} and \eqref{eq:linear_eq2} in matrix form
\begin{equation}
\begin{pmatrix}
1+\alpha^2\xi^2 & r(1+\alpha^2\xi^2)-\alpha(3+\alpha^2\xi^2) \\
\tan(\xi r)+\alpha\xi & r(\tan(\xi r)+\alpha\xi)-\alpha(2\tan(\xi r)+\alpha\xi) 
\end{pmatrix}
\begin{pmatrix}
C_1 \\
C_2
\end{pmatrix} = 0.
\end{equation}
For a non-trivial solution, we require 
\begin{align}
\det
\begin{pmatrix}
1+\alpha^2\xi^2 & r(1+\alpha^2\xi^2)-\alpha(3+\alpha^2\xi^2) \\
\tan(\xi r)+\alpha\xi & r(\tan(\xi r)+\alpha\xi)-\alpha(2\tan(\xi r)+\alpha\xi) 
\end{pmatrix} &= 0, \\
(1+\alpha^2\xi^2)(2\tan(\xi r)+\alpha\xi) - (3+\alpha^2\xi^2)(\tan(\xi r)+\alpha\xi) &= 0, \\
-\tan(\xi r) - 2\alpha\xi + \alpha^2\xi^2\tan(\xi r) &= 0.
\end{align}
So the root finding condition for $r$ is
\begin{equation}
    \tan(\xi r) = -\frac{2\alpha\xi}{1-\alpha^2\xi^2}.
\end{equation}

\bibliographystyle{unsrt}

\end{document}